\documentclass[prb,nofootinbib,twocolumn,eqsecnum,floatfix,endnotes]{revtex4}
\usepackage{epsfig}
\usepackage{amsmath}
\usepackage{amssymb}
\usepackage{mathrsfs}
\usepackage{color}
\usepackage{subfigure}
\usepackage{url}
\usepackage{enumerate}
\usepackage{braket}
\usepackage{comment}
\usepackage{yfonts}
\usepackage{enumerate}
\newcommand{\tr}[2]{\mathrm{Tr}_{#1}\left[#2\right]}
\newcommand{\btr}[2]{\mathrm{Tr}_{#1}\bigl[#2\bigr]}

\newcommand{\be}{\begin{equation}}
\newcommand{\ee}{\end{equation}}
\newcommand{\ba}{\begin{aligned}}
\newcommand{\ea}{\end{aligned}}

\newcommand{\bea}{\begin{eqnarray}}
\newcommand{\eea}{\end{eqnarray}}

\def\nn{\nonumber\\}
\def\fr#1{(\ref{#1})}

\newcommand{\ontop}[2]{\genfrac{}{}{0pt}{}{#1}{#2}}

\definecolor{dgreen}{RGB}{0,127,0}

\begin{document}
\title{Reduced Density Matrix after a Quantum Quench}
\author{Maurizio Fagotti}
\affiliation{\mbox{The Rudolf Peierls Centre for Theoretical Physics,
    Oxford University, Oxford, OX1 3NP, United Kingdom}}
\author{Fabian H.L. Essler}
\affiliation{\mbox{The Rudolf Peierls Centre for Theoretical Physics,
    Oxford University, Oxford, OX1 3NP, United Kingdom}}
\date{\today}
\begin{abstract}
We consider the reduced density matrix (RDM) $\rho_A(t)$ for a finite
subsystem $A$ after a global quantum quench in the infinite
transverse-field Ising chain. It has been recently shown that the
infinite time limit of $\rho_A(t)$ is described by the RDM
$\rho_{{\rm GGE},A}$ of a generalized Gibbs ensemble. Here we present some
details on how to construct this ensemble in terms of \emph{local}
integrals of motion, and show its equivalence to the expression in
terms of mode occupation numbers widely used in the literature.
We then address
the question, how $\rho_A(t)$ approaches $\rho_{{\rm GGE},A}$ as a
function of time. To that end we introduce a distance on
the space of density matrices and show that it approaches zero as a
universal power-law $t^{-3/2}$ in time. As the RDM completely
determines all local observables within $A$, this provides 
information on the relaxation of correlation functions of
local operators. We then address the issue, of how well a truncated
generalized Gibbs ensemble with a finite number of local higher
conservation laws describes a given subsystem at late times. We find
that taking into account only local conservation laws with a range at most
comparable to the subsystem size provides a good description. However,
excluding even a single one of the most local conservation laws in
general completely spoils this agreement.
\end{abstract}
\maketitle

%%%%%%%%%%%%%%%%%%%%%%%%%%
\section{Introduction}
%%%%%%%%%%%%%%%%%%%%%%%%%%
Recent advances in systems of optically trapped ultra-cold atomic
gases have made it possible to observe the nonequilibrium time
evolution of isolated many particle systems over long time scales
\cite{uc,kww-06,tc-07,tetal-11,cetal-12,getal-11}. A key property of
such cold atomic clouds is their weak coupling to the environment and
resulting smallness of external dissipative processes. To a good
approximation one is dealing with isolated quantum mechanical many
particle systems, which are prepared in generally mixed states, and
one is interested in the time dependence of observables, in particular at
late times. The experimental results have stimulated theoretical
efforts aimed at understanding the principles underlying the
nonequilibrium dynamics of isolated many particle systems. Some of the
most basic questions are whether observables generally relax to
time-independent values, and if they do, whether their stationary
values are described by a statistical ensemble. Other prominent issues
concern the roles of dimensionality and conservations laws.
Experiments on trapped ${}^{87}{\rm Rb}$ atoms\cite{kww-06} 
established that three-dimensional condensates rapidly relax to a
stationary state characterized by an effective temperature, whereas
constraining the motion of atoms to one dimension leads to a much
slower relaxation to a non-thermal distribution. It was argued that
this observed difference has its origin in the presence of additional
(approximate) conservation laws, related to quantum integrability, in
the one dimensional case. Theoretical efforts aimed at understanding
these and related questions
\cite{rev,gg,rdo-08,cc-07,2007_Gritsev_PRL_99,caz-06,bs-08,rsms-08,silva08,mwnm-09,2009_Moeckel_AP_324,fm-10,bkl-10,bhc-10,gce-10,2010_Mossel_NJP_12,2010_Barmettler_NJP_12,CEF:2011,rf-11,sfm-11,cic-11,2011_Mitra_PRL_107,igloi2011,
SE:2012,CEF1:2012,CEF2:2012,2012_Mossel_NJP_14,2012_Caux_PRL_109,2012_Foini_JSTAT_P09011,EEF:2012,BDKM:2012,MS:2012,GR:2012,KRSM:2012,RSM:2012,heyl,DT:2012,CDEO:2008,CE:2010}
indicate that in translationally invariant models there are at least
two basic types of behaviours at late times: subsystems either
thermalize~\cite{ETH}, i.e. are characterized by a Gibbs distribution with an
effective temperature, or they are described by a generalized Gibbs
ensemble (GGE) \cite{gg}. There is evidence that the latter applies to
integrable models, while the former constitutes the ``generic''
situation. 

A popular protocol for analyzing nonequilibrium evolution is a
so-called \emph{quantum quench}: here the system is originally
prepared in the ground state $|\Psi_0\rangle$ of some local, short
ranged Hamiltonian $H(h_0)$, where $h_0$ is a system parameter such as
a magnetic field or an interaction strength. At time $t=0$, $h_0$ is
then suddenly ``quenched'' to $h$, and the subsequent time evolution
under the new Hamiltonian $H(h)$ is studied. Under this protocol the
system remains in a pure state
$|\Psi_t\rangle=\exp(-iH(h)t)|\Psi_0\rangle$ at all times, and as a
whole can clearly never be described by a Gibbs or generalized Gibbs
distribution. This can be seen by considering the hermitian operators
\be
{\cal O}^{(n,m)}=|n\rangle\langle m|+|m\rangle\langle n|,
\ee
where $|n\rangle$ and $|m\rangle$ are eigenstates of $H(h)$ with
energies $E_n$ and $E_m$ respectively. Then the expectation values
\be
\langle\Psi_t|{\cal O}^{(n,m)}|\Psi_t\rangle=
\langle\Psi_0|n\rangle\langle m|\Psi_0\rangle e^{i(E_n-E_m)t}+{\rm h.c.}
\ee
are oscillating in time and never become stationary.
A useful and intuitive point of view is to focus on
\emph{local} properties of a given system in the thermodynamic limit,
i.e. ask questions only about observables contained in a finite
subsystem $A$ \cite{bs-08,CEF1:2012,CEF2:2012}. Here the (infinitely
large) complement $\bar{A}$ of $A$ can act as an effective bath, and
probability may freely dissipate from $A$ to $\bar{A}$. As a result
$A$ may be described by a mixed state. Arguably the most precise and
convenient description of this situation is in terms of the reduced
density matrix $\rho_A(t)$ of subsystem $A$. The latter is obtained
from the density matrix $\rho(t)=|\Psi_t\rangle\langle\Psi_t|$ of the
entire system as
\be
\rho_A(t)={\rm Tr}_{\bar{A}}\left[\rho(t)\right].
\ee
A central question is then, whether for any finite subsystem $A$
\be
\lim_{t\to\infty}\rho_A(t)=\rho_{{\rm stat},A},
\label{stat_state}
\ee
where $\rho_{{\rm stat},A}$ is a time-independent reduced density
matrix obtained as
\be
\rho_{{\rm stat},A}={\rm Tr}_{\bar{A}}\left[\rho_{\rm stat}\right].
\ee
If \fr{stat_state} holds, then the system evolves towards a stationary
state described by the distribution $\rho_{\rm stat}$. In particular
\fr{stat_state} implies that the expectation values of \emph{any}
local operator ${\cal O}_A$ acting only within subsystem $A$ is
given by 
\be
\lim_{t\to\infty}\langle\Psi_t|{\cal O}_A|\Psi_t\rangle=
{\rm Tr}\left[\rho_{\rm stat}\ {\cal O}_A\right].
\ee
An efficient way of investigating whether a given RDM approaches a
known stationary distribution at late times  was introduced in
Ref.~[\onlinecite{bhc-10}] by considering the operator norm $\parallel
\rho_A(t)-\rho_{{\rm stat},A}\parallel_{\rm op}$. If this approaches
zero at late times, then the system relaxes locally to the stationary
distribution $\rho_{\rm stat}$. Ref.~[\onlinecite{bhc-10}] was
concerned with the case where $\rho_{\rm stat}$ describes a thermal
ensemble with a given effective temperature, and considered very small
subsystems. Here we are interested in a quench to a quantum integrable
model. As we have alluded to before, the stationary state for such
quenches is believed to be described locally by a generalized Gibbs
ensemble (for the model we consider below this was established in
Ref.~[\onlinecite{CEF2:2012}]). More precisely, the density matrix of
the entire system is expected to be of the form
\be
\rho_{\rm stat}=\rho_{\rm
  GGE}=\frac{1}{\cal Z}e^{\sum_n\lambda_n I_n}\ ,
\label{rho_GGE}
\ee
where $Z$ is a normalization\footnote{We have in mind regularizing the
  system by enclosing it in a very large but finite volume.}, and
$I_n$ are \emph{local} conserved quantities, i.e. local operators such
that
\be
[I_n,I_m]=0=[I_m,H(h)].
\ee
The Lagrange mutipliers $\lambda_n$ are fixed by the requirements
\be
\langle\Psi_0|I_n|\Psi_0\rangle={\rm Tr}\left[\rho_{\rm GGE}I_n\right].
\ee
We stress that the GGEs considered in the quench context are
fundamentally different from thermal ensembles, because through the
specific values of the Lagrange multipliers they retain an infinite
amount of information about the initial state. Above we have
stipulated that only local (in space) conservation laws $I_n$ are to
be included in the definition of $\rho_{\rm GGE}$, but it is in fact a
matter on ongoing debate whether locality is a necessary or even
desirable requirement \cite{nonlocal}. In this context a result
obtained in Ref.~[\onlinecite{CEF1:2012}] is rather illuminating:
there it was demonstrated for a particular example, the transverse
field Ising chain, that different statistical ensembles can have
identical local properties. The two ensembles considered were a GGE of
the form \fr{rho_GGE}, and the so-called ``pair ensemble'' obtained
by time averaging the quench density matrix $\rho(t)$. Given that
$\rho_{\rm stat}$ is generally not unique, it is clearly desirable to
identify the simplest description. To that end we introduce
\emph{truncated generalized Gibbs ensembles} of the form
\be
\rho^{(n_0)}_{\rm tGGE}=\frac{1}{\cal Z}e^{\sum_{n<n_0}\lambda_n I_n}\ ,
\ee
and investigate how well such ensembles describe the stationary state
for quenches to integrable models.

The outline of this paper is as follows. In section \ref{sec:facts} we
review some relevant results for the transverse field Ising
chain. Local conservation laws are presented in section
\ref{s:locclaws} and used in sections \ref{sec:GGE}, \ref{sec:PGGE},
\ref{sec:DGGE} to define several classes of generalized Gibbs
ensembles. Properties of corresponding reduced density matrices are
discussed in section \ref{ss:denmat}. In section \ref{s:distance} we
discuss general properties of distances on the space of reduced
density matrices and  introduce the distance used in the
remainder of the paper. In sections \ref{sec:1spind}, \ref{sec:larger},
\ref{s:SSB} and \ref{sec:across} we present results for the distance
between quench and generalized Gibbs reduced density matrices. We
summarize our results in section \ref{sec:summary}. Various technical
issues are discussed in four appendices.
%%%%%%%%%%%%%%%%%%%%%%%%%%%%%%%%%%%%%%%%%%%%%%%%%%%%
\section{Some Facts about the Transverse-Field Ising Chain (TFIC)}
\label{sec:facts}
%%%%%%%%%%%%%%%%%%%%%%%%%%%%%%%%%%%%%%%%%%%%%%%%%%%%
Here we briefly review some relevant results on the TFIC. The latter
is an important paradigm for quantum phase transitions in equilibrium
\cite{sachdev} as well as non-equilibrium dynamics\cite{bdm-70,rsms-08,igloi2011,
CEF:2011,CEF1:2012,CEF2:2012,EEF:2012}. 
In the latter context experimental realizations range from
cold atomic gases \cite{trotzky} to circuit QED \cite{VDM:2013}.
The Hamiltonian of the model on a ring is
\be\label{eq:HIs}
H(h)=-J\sum_{j=1}^L\Bigl[\sigma_{j}^x\sigma_{j+1}^x+h\sigma_k^z\Bigr],
\ee
where $\sigma^\alpha_{L+1}=\sigma^\alpha_1$. The quantum spins
can be mapped to (real) Majorana fermions by means of the
Jordan-Wigner transformation 
\be\label{eq:JW}
a_{2\ell}=\left(\prod_{j=1}^{\ell-1}\sigma_j^z\right)\sigma_\ell^y\,
,\qquad 
a_{2\ell-1}=\left(\prod_{j=1}^{\ell-1}\sigma_j^z\right)\sigma_\ell^x\, ,
\ee
where $\{a_i,a_j\}=2\delta_{i j}$. In terms of the Majorana
fermions~\eqref{eq:JW}, the Hamiltonian takes a block-diagonal form
\bea\label{eq:Hf}
H(h)&=&\frac{1+e^{i\pi \mathcal N}}{2}H_{\rm R}+\frac{1-e^{i\pi \mathcal
    N}}{2}H_{\rm NS}\ ,\nn
H_{\rm NS/R}&=&iJ\sum_{j=1}^{L-1}a_{2j}\left[a_{2j+1}-h a_{2j-1}\right]\nn
&&-iJa_{2L}\left[ha_{2L-1}\mp a_{1}\right].
\eea
Here $\mathcal N$ is the number operator 
\be\label{eq:calN}
\mathcal N=\sum_{j=1}^L\frac{\sigma^z_j-1}{2}
=\sum_{j=1}^L\frac{ia_{2j}a_{2j-1}-1}{2},
\ee
and by construction $e^{i \pi\mathcal N}=\prod_j\sigma_j^z$ commutes
with $H_{\rm R,NS}$. The two blocks $H_{\rm R}$ and $H_{\rm NS}$
correspond to periodic and antiperiodic boundary conditions on the
fermions respectively. They can be diagonalized by Bogoliubov transformations
\bea
a_{2j-1}&=&\frac{1}{\sqrt{L}}\sum_pe^{i\frac{\theta_p}{2}-ipj}
\left[\alpha_p+\alpha_{-p}^\dagger\right] ,\nn
a_{2j}&=&-\frac{i}{\sqrt{L}}\sum_pe^{-i\frac{\theta_p}{2}-ipj}
\left[\alpha_p-\alpha_{-p}^\dagger\right],
\label{bogo}
\eea
where the Bogoliubov angle $\theta_p$ is given by
\be
e^{i\theta_p}=\frac{h-e^{ip}}{\sqrt{1+h^2-2h\cos p}}.
\label{theta_k}
\ee
The diagonal form of the Hamiltonian is
\be
H_{\rm NS}(h)=\sum_{p\in{\rm NS}}
\varepsilon_h(p) \left(\alpha^\dagger_p\alpha_p-\frac{1}{2}\right),
\ee
where the single-particle energy is given by
\be
\varepsilon_h(k)=2J\sqrt{1+h^2-2h\cos k}.
\label{veps}
\ee
The ground states of $H_{\rm R,NS}(h)$ are the fermionic vacua
\be
\alpha_p|0;h\rangle_{\rm R,NS}=0\ .
\ee
Here the momenta are $p=\frac{\pi n}{L}$, where $n$ are even/odd integers
for R and NS fermions respectively.
%%%%%%%%%%%%%%%%%%%%%%%%%%%%%%
\subsection{Quantum Quenches}
%%%%%%%%%%%%%%%%%%%%%%%%%%%%%%
Our quench protocol is as follows: we prepare the system in the
ground state $|\Psi_0\rangle$ for an initial value $h_0$ of the
transverse magnetic field. At time $t=0$ we instantaneously change the
field from $h_0$ to $h$. The state of the system at times $t>0$ is
obtained by evolving with respect to the new Hamiltonian $H(h)$
\be
|\Psi_t\rangle=e^{iH(h)t}|\Psi_0\rangle.
\ee
An important quantity is the difference $\Delta_k=\theta_k-\theta_k^0$
of Bogolibov angles required to diagonalize $H(h)$ and $H(h_0)$ respectively
\bea
\cos\Delta_k&=&\frac{4J^2(1+h h_0 -(h+h_0) \cos k)}{
\varepsilon_h(k)\varepsilon_{h_0}(k)}.
%,\nn
%\sin\Delta_k&=&\frac{(2J)^2(h-h_0)\sin k}{
%\varepsilon_h(k)\varepsilon_{h_0}(k)},\nn
\label{cosdeltak}
\eea
As we are interested in obtaining results in the thermodynamic limit
we have to distinguish between two cases.
%%%%%%%%%%%%%%%%%%%%%%%%%%%%%%
\subsubsection{Quenches from the Paramagnetic Phase $h_0>1$}
%%%%%%%%%%%%%%%%%%%%%%%%%%%%%%
Here the initial state in a large, finite volume is simply the NS
vacuum
\be
|\Psi_0\rangle=|0;h_0\rangle_{\rm NS}.
\ee
The time evolved state can then be written in the form \cite{CEF1:2012}
\be
\label{Psit_para}
|\Psi_t\rangle=\frac{1}{\cal M}\exp\Bigl[i {\displaystyle\sum_{0<p\in
    {\rm NS}}} \tan\big(\frac{\Delta_p}{2}\big) e^{-2i\varepsilon_p t}\alpha_{-p}^\dag
\alpha_p^\dag\Bigr]\ket{0;h}_{\rm NS},
\ee
where $|0;h\rangle_{\rm NS}$ is the ground state of $H_{\rm NS}(h)$
and ${ \cal M}$ a normalization factor.

%%%%%%%%%%%%%%%%%%%%%%%%%%%%%%
\subsubsection{Quenches from the Ferromagnetic Phase $h_0<1$}
%%%%%%%%%%%%%%%%%%%%%%%%%%%%%%
Here our initial state in a large, finite volume must reflect the
spontaneous symmetry breaking of the $\mathbb{Z}_2$ spin-flip symmetry
$\sigma^{x,y}\rightarrow-\sigma^{x,y}$ in the thermodynamic limit. The
appropriate choice is \cite{CEF1:2012}
\be
\ket{\Psi_0}=\frac{|0;h_0\rangle_{\rm R}+|0;h_0\rangle_{\rm NS}}{\sqrt{2}}\, .
\label{ferroGS}
\ee

%%%%%%%%%%%%%%%%%%%%%%%%%%%%%%%%%%%%%%%%%%%%%%%%%
\section{Local Conservation Laws in the TFIC}\label{s:locclaws}  %
%%%%%%%%%%%%%%%%%%%%%%%%%%%%%%%%%%%%%%%%%%%%%%%%%%

We consider the one dimensional transverse field Ising chain in the
thermodynamic limit
\be
H=-J\sum_{n=-\infty}^\infty \sigma^x_n\sigma^x_{n+1}+h\sigma^z_n.
\ee
Following Ref.~[\onlinecite{prosen}] we can construct an infinite
number of \emph{local} conservation laws $I_n^\pm$ 
\be
[I_n^\alpha,I_m^\beta]=0\ ,\quad n=0,1,\dots, \alpha,\beta=\pm,
\ee
where the Hamiltonian itself is $H=I_0^+$.
Let us define operators
\bea
U_{n>0}&=&\frac{1}{2}\sum_{j=-\infty}^\infty
\sigma^x_j\left(\prod_{l=1}^{n-1}\sigma^z_{j+l}\right)\sigma^x_{j+n}\ ,\nn
U_0&=&-\frac{1}{2}\sum_{j=-\infty}^\infty\sigma^z_j\ ,\nn
U_{n<0}&=&\frac{1}{2}\sum_{j=-\infty}^\infty
\sigma^y_j\left(\prod_{l=1}^{|n|-1}\sigma^z_{j+l}\right)\sigma^y_{j+|n|},
\eea
and
\bea
V_{n>0}&=&\frac{1}{2}\sum_{j=-\infty}^\infty
\sigma^x_j\left(\prod_{l=1}^{n-1}\sigma^z_{j+l}\right)\sigma^y_{j+n}\ ,\nn
V_{n<0}&=&-\frac{1}{2}\sum_{j=-\infty}^\infty
\sigma^y_j\left(\prod_{l=1}^{|n|-1}\sigma^z_{j+l}\right)\sigma^x_{j+|n|}.
\eea
In terms of these operators the local conservation laws are
\bea
I^+_n&=&-J(U_{n+1}+U_{1-n})+hJ (U_n+U_{-n})\ ,\nn
I^-_n&=&J(V_{n+1}+V_{-n-1})\ ,\quad n\geq 0.
\eea
They are local in the sense that the density of $I_n^\alpha$ involves
only spins on $n+2$ neighbouring sites. By virtue of their locality,
the conservation laws can all be expressed in terms of Jordan-Wigner
fermions \fr{eq:JW}
\bea
I^+_n&=&
\frac{i}{2}\sum_{j=-\infty}^\infty Ja_{2j}[a_{2j+2n+1}+a_{2j-2n+1}],\nn
&&\qquad\quad-hJa_{2j}[a_{2j+2n-1}+a_{2j-2n-1}]\ ,\nn
I^-_{n-1}&=&-\frac{iJ}{2}\sum_{j=-\infty}^\infty a_{2j}a_{2j+2n}
+a_{2j-1}a_{2j+2n-1} .\
\label{Ipm}
\eea
We now realize that all conservation laws \eqref{Ipm} are in fact
quadratic in Majorana fermions! It is then a simple matter to
diagonalize them simultaneously by means of a Bogoliubov
transformation \fr{bogo}, where on the infinite chain the Bogoliubov
fermion operators have anticommutation relations 
\be
\{\alpha_p,\alpha^\dagger_k\}=2\pi\delta(p-k).
\ee
The conservation laws take the simple form
\be
\ba
I_n^+&=\int_{-\pi}^\pi\frac{dp}{2\pi}\ \cos(nk)\varepsilon(k)
\alpha^\dagger_k\alpha_k\ ,\\  
I_{n}^-&=-\int_{-\pi}^\pi\frac{dp}{2\pi}\ 
2J\sin\big((n+1)k\big)\alpha^\dagger_k\alpha_k\ ,
\label{ipm_2}
\ea
\ee
which furthermore shows that they are even/odd under spatial reflections.
Interestingly the conservation laws $I_n^-$ do not depend on the
transverse field $h$ and are therefore \emph{shared} by the entire
one-parameter family of Hamiltonians $H(h)$. This seems to be a
generic feature of models with a free fermion spectrum like the TFIC,
see App.~\ref{a:cl} and Ref.~[\onlinecite{F2:2012}].

%%%%%%%%%%%%%%%%%%%%%%%%%%%%%%%%%%%%%%%%%%%%
\subsection{Local conservation laws for periodic boundary conditions}
%%%%%%%%%%%%%%%%%%%%%%%%%%%%%%%%%%%%%%%%%%%%
Above we focussed on the bulk contribution to the local conservation
laws. For a finite system on a ring there are boundary contributions,
which can be determined as follows. In terms of the Bogoliubov
fermions, the conservation laws for periodic boundary conditions are
\bea
I_n^+&=&\sum_k \cos(nk)\varepsilon(k)
\alpha^\dagger_k\alpha_k\ ,\\  
I_{n}^-&=&-\sum_k\ 
2J\sin\big((n+1)k\big)\alpha^\dagger_k\alpha_k\ ,
\label{ipm_2_pbc}
\eea
where the momenta are taken to be either in the R or NS sectors. By
inverting the Bogoliubov transformation and Fourier transforming back
to position space, one obtains a representation of the conservation
laws in terms of the Majorana fermions $a_j$ for periodic/antiperiodic
boundary conditions respectively. Finally, inverting the Jordan-Wigner
transformation gives the desired expression in terms of spins. 

%%%%%%%%%%%%%%%%%%%%%%%%%%%%
\section{Generalized Gibbs Ensemble}
\label{sec:GGE}
%%%%%%%%%%%%%%%%%%%%%%%%%%%%
We now \emph{define} the density matrix of a generalized Gibbs
ensemble formally by the expression
\be
\rho_{\rm GGE}=\frac{1}{\cal
  Z}\exp\left(-\sum_{n=0}^\infty\sum_{\sigma=\pm}
\left[\lambda^\sigma_nI_n^\sigma\right]\right),
\label{rhoGGE}
\ee
where ${\cal Z}$ is a normalization that ensures ${\rm Tr}\rho_{\rm GGE}=1$.
In practice we need to regularize \fr{rhoGGE} in an asymptotically large,
finite volume $L$. 

\begin{figure}[t]
\includegraphics[width=0.47\textwidth]{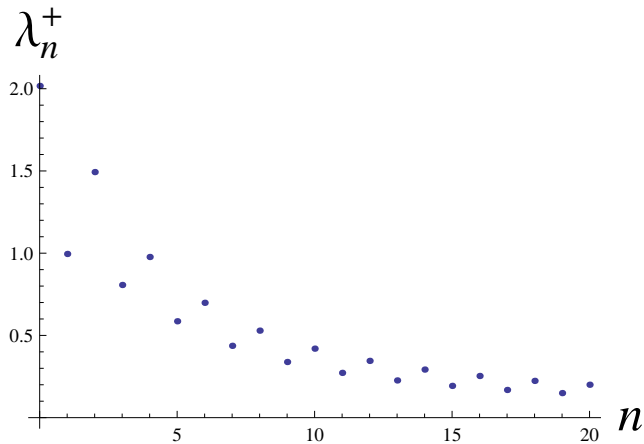}
\caption{Parameters $\lambda_n$ for a quench within the ordered phase
from $h_0=0.1$ to $h=0.7$.}
\label{fig:lambda}
\end{figure}

The Lagrange multipliers $\lambda_n^\sigma$ are fixed through the
requirements
\be
\lim_{L\to\infty}\frac{1}{L}{\rm Tr}\left[\rho_{\rm
    GGE}I_n^\sigma\right]=
\lim_{L\to\infty}\frac{\langle\Psi_0|I_n^\sigma|\Psi_0\rangle}{L}.
\label{lambdas}
\ee
Using translational invariance we can alternatively work with the
densities of the conservation laws 
\be
I_n^\sigma=\sum_{j=-\infty}^\infty
(I_n^\sigma)_{j,\dots,j+n+1}\ ,
\label{densities}
\ee
to rewrite \fr{lambdas} as
\be
{\rm Tr}\left[\rho_{\rm GGE}(I_n^\sigma)_{j,\dots,j+n+1}\right]
=
\langle\Psi_0|(I_n^\sigma)_{j,\dots,j+n+1}|\Psi_0\rangle.
\label{lambdas2}
\ee
The solution to this system of equations is
\bea
\lambda^+_l&=&\Bigl({2-\delta_{l,0}}\Bigr)\int_{-\pi}^\pi\frac{dk}{\pi}
\frac{\cos(lk)}{\varepsilon_h(k)}  
\mathrm{arctanh}(\cos\Delta_k)\ ,\nn 
\lambda^-_l&=&0\ ,
%-\int_{-\pi}^\pi\frac{dk}{\pi}\frac{\sin\big((l+1)k\big)}{2J} \gamma_k\ ,
\label{lagr}
\eea
where $l\geq 0$ and $\cos\Delta_k$ is defined in \fr{cosdeltak}. In
Fig.~\ref{fig:lambda}  we show $\lambda_l^+$ for a quench from
$h_0=0.1$ to $h=0.7$. 
The large $l$ behaviour of Eq.~\eqref{lagr} is determined by the
regions $k\sim0,\pi$ (where the integrand has a logarithmic
singularity) and one can show that
\be
\lambda_l^+\sim\frac{2}{l}\Bigl(\pm\frac{1}{\varepsilon_h(0)}
+\frac{(-1)^l}{\varepsilon_h(\pi)}\Bigr), 
\ee
where the sign is $+$ for quenches within the same phase and $-$ otherwise.
We see that the Lagrange multipliers $\lambda_n^+$ decay rather slowly
as a function of $n$.

%%%%%%%%%%%%%%%%%%%%%%%%%%%%%%%%%%%%%%%%%%%%%%%%%%%%%
\subsection{GGE in terms of mode occupation numbers}
%%%%%%%%%%%%%%%%%%%%%%%%%%%%%%%%%%%%%%%%%%%%%%%%%%%%%

In the literature the generalized Gibbs ensemble is
often constructed from mode occupation numbers
$n_k=\alpha^\dagger_k\alpha_k$, see e.g.
\cite{gg,caz-06,fm-10,lm-10,CCRSS:2011}.
The latter are non-local (in space) as they involve a Fourier transform.
We will now establish the relation between this and our definition
\fr{rhoGGE}. It follows from \eqref{ipm_2} that the density matrix
can be rewritten in the form 
\be
\rho_{\rm GGE}=\frac{1}{\cal Z}\exp\left(-\int_{-\pi}^\pi\frac{dk}{2\pi}
\gamma_k\alpha^\dagger_k\alpha_k\right)\ ,
\label{rhoGGE2}
\ee
where
\be
\gamma_k=\sum_{n=0}^\infty \lambda^+_n \varepsilon_h(k)\cos(kn)
-2J\lambda^-_n\sin\big(k(n+1)\big).
\label{gammak}
\ee
This establishes the fact that the GGE density matrix can be
constructed either from the local conservation laws \eqref{Ipm}, or from
the mode occupation numbers $n_k$. This relationship generalizes to
interacting integrable models, where the appropriate GGE can be
formulated either in terms of the local integrals of motion generated
from the transfer matrix, or from the mode occupation numbers 
$n_a(k)=Z^\dagger_a(k)Z_a(k)$, where $Z_a(k)$ are
Faddeev-Zamolodchikov operators \cite{review,Mussardo}.

%%%%%%%%%%%%%%%%%%%%%%%%%%%%%%%%%%%%%%%%%%%%%%%%%%%%
\section{Truncated Generalized Gibbs Ensembles}
\label{sec:PGGE}
%%%%%%%%%%%%%%%%%%%%%%%%%%%%%%%%%%%%%%%%%%%%%%%%%%%%%
In order to assess the importance of the various conserved quantities,
it is useful to define ensembles that interpolate between the Gibbs
distribution and the GGE. We define particular such truncated GGEs as
follows. Given that the densities of the conservation laws $I_n^\pm$
involve $n+2$ neighbouring sites, it is natural to retain only
the ``most local'' conservation laws, i.e.
\be
\rho_{\rm tGGE}^{(y)}=\frac{1}{\mathcal Z_y}\exp\Bigl(\sum_{n=0}^{y-1}\sum_{\sigma=\pm}[\lambda_{n,y}^\sigma I_n^\sigma]\Bigr)\, .
\label{rho_tGGE}
\ee
Here $y$ is an integer and $y=1$ ($y=\infty$) corresponds to the Gibbs
ensemble (GGE). The Lagrange multipliers $\lambda_{n,y}^\sigma$ are
obtained from the requirements
\be\label{eq:tGGEreq}
\btr{}{(I_n^\sigma)_{j,\dots,j+n+1}\ \rho_{\rm tGGE}^{(y)}}=
\braket{\Psi_0|(I_n^\sigma)_{j,\dots,j+n+1}|\Psi_0},
\ee
where $0\leq n<y$. Eqns \fr{eq:tGGEreq} are a consequence of
$[I^\sigma_n,H]=0$ and the assumption that the stationary state after
the quench is described by RDMs based on \fr{rho_tGGE}. 
For transverse field quenches we have
$\lambda_{n,y}^-=0$, but the other Lagrange multipliers are different
from their respective values in the full GGE, i.e.
\be
\lambda_{n,y}^+\neq \lambda_{n}^+.
\ee
We note that the correlation matrix of $\rho^{(y)}$ can be computed
efficiently using FFT algorithms. This is in contrast to the case of
theories with non-trivial scattering matrices, for which it is
extremely difficult to reconstruct the Lagrange multipliers from the
conservation laws \cite{Mpc}.

%%%%%%%%%%%%%%%%%%%%%%%%%%
\section{Defective Generalized Gibbs Ensembles}%
\label{sec:DGGE}                                                   %
%%%%%%%%%%%%%%%%%%%%%%%%%%
It is instructive to consider a second type of truncated GGE, where we
retain an infinite, but incomplete set of integrals of motion. Such
``defective'' GGEs will allow us to ascertain the role of a
particular local conservation law. We define the truncated defective GGE as the density matrix ($q<y$)
\be
\rho_{\rm tdGGE}^{(+q),y}=\frac{1}{\mathcal
  Z_{(+q),y}}\exp\Bigl(\sum_{\ontop{n=0}{n\neq q}}^{y}
[\lambda_{n,(+q),y}^+ I_n^+]\Bigr)\, , 
\label{rho_dtGGE}
\ee
in which the Lagrange multipliers $\lambda_{n,(+q)}^+$ are fixed by the system \fr{lambdas} with $n\leq y$, $n\neq q$, %the equation $(n,\sigma)=(q,+)$ has
%been dropped 
and we have used that the Lagrange multipliers
$\lambda_{n,(+q),y}^-$ must vanish as a consequence of reflection
symmetry around the origin.
We then define the defective GGE as the limit $y\rightarrow\infty$ of truncated defective GGEs:
\be\label{rho_dGGE}
\rho_{\rm dGGE}^{(+q)}=\lim_{y\rightarrow\infty}\rho_{\rm tdGGE}^{(+q),y}\, .
\ee
%where $\rho_{\rm tdGGE}^{(+q),y}$ is a truncated defective GGE in which the $y$ most local integrals of motion 
%defined by ($q<y$)
%\be
%\rho_{\rm tdGGE}^{(+q),y}=\frac{1}{\mathcal
%  Z_{(+q),y}}\exp\Bigl(\sum_{\ontop{n=0}{n\neq q}}^{y}
%[\lambda_{n,(+q),y}^+ I_n^+]\Bigr)\, , 
%\label{rho_dtGGE}
%\ee
%where we have used that the Lagrange multipliers
%$\lambda_{n,(+q),y}^-$ must vanish as a consequence of reflection
%symmetry around the origin.
%{\color{blue}in which } the Lagrange multipliers $\lambda_{n,(+q)}^\sigma$ are fixed by the
%system \fr{lambdas}, where the equation with $(n,\sigma)=(q,+)$ has
%been dropped. 
In order to solve the system of equations \fr{lambdas} for the defective GGE it
is useful to work in the mode occupation number representation
\fr{rhoGGE2}, which reads
\be
\rho_{\rm dGGE}^{(+q)}=\frac{1}{\cal
  Z}_{(+q)}\exp\left(-\int_{-\pi}^\pi\frac{dk}{2\pi} 
\gamma_k^{(+q)}\alpha^\dagger_k\alpha_k\right)\ ,
\label{rhodGGE2}
\ee
where 
%\bea
%\gamma_k^{(+q)}&=&\sum_{n=0}^\infty \lambda^+_{n,(+q)}
% \varepsilon_h(k)\cos(kn)\nn
%&&\qquad-2J\lambda^-_{n,(+q)}\sin\big(k(n+1)\big).
%\label{gammak2}
%\eea
%The inverse relations are
%\bea
%\lambda^+_{l,(+q)}
%&=&\Bigl({1-\frac{\delta_{l 0}}{2}}\Bigr)\int_{-\pi}^\pi\frac{dk}{\pi}
%\frac{\cos(lk)}{\varepsilon_h(k)}\gamma_k^{(+q)}\ ,\nn
%\lambda^-_{l,(+q)}&=&-\int_{-\pi}^\pi\frac{dk}{\pi}\frac{\sin\big((l+1)k\big)}{2J} 
%\gamma_k^{(+q)}\ .
%\label{lagr_2}
%\eea
the Lagrange multipliers $\gamma_k^{(+q)}$ are subject to the set of
equations
\bea\label{eq:constraints}
\int_{-\pi}^{\pi}\frac{\mathrm d
  k}{2\pi}\Bigl[\tanh\Bigl(\frac{\gamma_k^{(+q)}}{2}\Bigr)-\cos\Delta_k\Bigr]\varepsilon(k)\cos(n
k)=0\ ,\nn
\int_{-\pi}^{\pi}\frac{\mathrm d k}{2\pi}\Bigl[\tanh\Bigl(\frac{\gamma_k^{(+q)}}{2}\Bigr)-\cos\Delta_k\Bigr]\sin((n+1) k)=0.
\eea
Guided by the fact that $\cos(nk)$ and $\sin((n+1)k)$ form an
orthonormal set of functions on $[-\pi,\pi]$, we look for a solution
of the form 
\be\label{eq:kappa}
\tanh\Bigl(\frac{\gamma_{k}^{(+q)}}{2}\Bigr)=\cos\Delta_k-\kappa_q^+ \frac{\cos(q k)}{\varepsilon(k)}\, ,
\ee
where $\kappa_q^+$ is a yet to be determined constant. We note that
the value of $\kappa^+_q$ affects the expectation values of local
operators. For some cases $\kappa^+_q$ can be easily determined as
follows. Given that $|\tanh(x)|\leq 1$, \fr{eq:kappa} implies that
\be\label{eq:bounds}
\Bigl|\cos\Delta_k-\kappa_q^+ \frac{\cos(q k)}{\varepsilon(k)}\Bigr|\leq 1\qquad \forall k\, .
\ee
Setting $k=0,\pi$ then gives
\be\label{eq:bounds1}
\ba
|2J (h-1)\mathrm{sgn}(h_0-1)-\kappa_q^+|&\leq 2 J |h-1|\\
|2J(h+1)-(-1)^q\kappa_q^+|&\leq 2J(h+1)\, .
\ea
\ee
Eqs~\eqref{eq:bounds1} allow us to identify cases, in which $\kappa_q^+=0$:
\begin{enumerate}
\item{} odd $q$ and quenches within the same phase; 
\item{} even $q$ and quenches across the critical point.
\end{enumerate}
Importantly, $\kappa_q^+=0$ implies that $\rho_{\rm
  dGGE}^{(+q)}\equiv\rho_{\rm GGE}$, i.e. the defective GGE is
identical to the full GGE. This ``GGE reconstruction'' is a
peculiarity of free-fermion models and can be traced back to the
existence of conservation laws independent of the quench parameter,
see also Ref.~[\onlinecite{F2:2012}]. 

For general quenches and values of $q$, $\kappa_q^+$ is determined by Eq.~\eqref{rho_dGGE}. We find that  it  takes the value corresponding to the maximal entanglement entropy  (as shown in Fig.~\ref{fig:local2}), although the entanglement entropy may be non-stationary under a variation of the excluded integral of motion (see Appendix~\ref{app:defective} for further details).
\section{Reduced Density Matrices in the Transverse Field Ising Chain}\label{ss:denmat} % 
%%%%%%%%%%%%%%%%%%%%%%%%%%%%%%%%%%%%%%%%%%%%%

In this section we summarize some basic features of RDMs in the
TFIC. We note that most of the following discussion generalizes straightforwardly
to other spin chains with free fermionic spectra such as the quantum
XY model. Our starting point is a density matrix $\rho$ describing the
entire system, which we take to be of size $L$ with periodic boundary
conditions. We are interested in the limit $L\to\infty$, but it is
convenient to start with a large, finite chain. The RDM of a subsystem
consisting of $\ell$ spins $1/2$ at sites $x_i$, $i=1,\ldots,\ell$ can
be expressed in the form 
\be\label{eq:RDMgen}
\rho_{\{x_1,\cdots,x_\ell\}}=\frac{1}{2^\ell}\sum_{\{\alpha\}_\ell}\tr{}{\rho\,
  \sigma_{x_1}^{\alpha_1}\cdots\sigma_{x_\ell}^{\alpha_\ell}}\sigma_{x_1}^{\alpha_1}\cdots\sigma_{x_\ell}^{\alpha_\ell}\,
, 
\ee
where $\alpha_i=0,x,y,z$ and $\sigma^{0}\equiv \mathrm I$. The quantum spins
are mapped to (real) Majorana fermions by the Jordan-Wigner transformation 
\fr{eq:JW}. The nonlocality of the transformation~\eqref{eq:JW} has
important consequences for RDMs. First and foremost, if the spins are not
adjacent, the map from spin to fermionic degrees of freedom does not
have a simple reduction to the subspace of the Hilbert space formed
by sites $\{x_1,\cdots,x_\ell\}$, because of Jordan-Wigner strings
stretching between sites~\cite{FC:2010,F:2012}. However, the RDM of a block of adjacent
spins can be mapped one-to-one on a block of adjacent fermions~\cite{vidal,pe-rev}, provided
that the first site of the block coincides with site $1$, i.e. the
origin of Jordan-Wigner strings. Then \fr{eq:RDMgen} can be
represented in the form 
\be\label{eq:RDMgenf}
\rho_\ell=\frac{1}{2^\ell}\sum_{\{\mu\}}\tr{}{\rho\, a_1^{\mu_1}\cdots a_{2\ell}^{\mu_{2\ell}}} a_{2\ell}^{\mu_{2\ell}}\cdots a_1^{\mu_1}
\ee
with $\mu_j=0,1$. An important quantity in what follows is the
\emph{correlation matrix} $\Gamma$
\be
\Gamma_{i j}=\tr{}{\rho\, a_j a_i}-\delta_{i j}\ ,\quad 1\leq i,j\leq 2\ell.
\ee
In the cases of interest to us, the correlation matrix is of
block-Toeplitz form
\be\label{eq:Gamma}
\Gamma = \left[
 \begin{array}{ccccc}
\mathtt\Gamma_{0}  & \mathtt\Gamma_{-1}   &   \cdots & \mathtt\Gamma_{1-\ell}  \\
\mathtt\Gamma_{1} & \mathtt\Gamma_{0}   & &\vdots\\
\vdots&  & \ddots&\vdots  \\
\mathtt\Gamma_{\ell-1}& \cdots  & \cdots  &\mathtt\Gamma_{0}
\end{array}
\right],
\ee
where 
\be
\label{eq:Gamma_l}
\Gamma_l=\int_{-\pi}^\pi\frac{dk}{2\pi}e^{-ilk}
\left(
\begin{array}{cc}
-f(k)&g(k)\\
-g(-k)&f(k)
\end{array}
\right).
\ee
The TFIC Hamiltonian exhibits a $\mathbb{Z}_2$ symmetry 
\be
a_j\longrightarrow -a_j\ .
\label{Z2}
\ee
If the density $\rho$ is invariant under the transformation \fr{Z2}
we have $\tr{}{\rho\, a_j}=0$, and as Wick's theorem applies to the
Jordan-Wigner fermions we can express  \eqref{eq:RDMgenf} as a
Gaussian~\cite{pe-rev}  
\be\label{eq:gaussian}
\rho_{\ell}=\frac{1}{Z}e^{\frac{1}{4}\sum_{m n}a_m W_{m n}a_n}\, ,
\ee
where $Z$ ensures that ${\rm Tr}\rho_\ell=1$ and $W$ is a skewsymmetric,
$2\ell$-by-$2\ell$ Hermitian matrix related to $\Gamma$ by
\be
\tanh\frac{W}{2}=\Gamma\ .
\label{WGamma}
\ee
We now turn to the three particular cases of interest, namely those where
$\rho$ in \eqref{eq:RDMgen} is a thermal density matrix, a GGE density
matrix, or the density matrix after a global quantum quench of the
transverse field in the TFIC.   

%%%%%%%%%%%%%%%%%%%%
\subsection{Thermal Density Matrix}  %
%%%%%%%%%%%%%%%%%%%%
On a very large ring, the Hamiltonian has a block diagonal structure,
see section \ref{sec:facts}. The thermal density matrix is a function of the
Hamiltonian and therefore inherits the same block structure 
\be
\rho_\beta=\Bigl[\frac{1+e^{i\pi \mathcal
      N}}{2}\frac{e^{-\beta H_{\rm R}}}{Z_{\rm R}}+
\frac{1-e^{i\pi \mathcal N}}{2}\frac{e^{-\beta  H_{\rm NS}}}{Z_{\rm NS}}\Bigr]\, .
\ee
It follows from this that only even operators have non-vanishing
expectation values, i.e.
\be
{\rm Tr}\left(\rho_\beta{\cal O}\right)\neq0\rightarrow [e^{i\pi{\cal
      N}},{\cal O}]=0.
\ee
In the thermodynamic limit the difference between expectation values
of local operators with respect to the R and NS sectors tends to zero,
so that we may work exclusively in e.g. the R sector. The resulting
RDM of a contiguous block of spins is then Gaussian
\eqref{eq:gaussian}, \fr{WGamma} with 
\be
\left(\Gamma_\beta\right)_{ij}
= \tr{}{\frac{e^{-\beta H_R}\, a_{j} a_i}{Z_R}}-\delta_{i j}\, .
\ee 
It can be written in the form \fr{eq:Gamma} with
\be
 \label{eq:thermal}
f(k)=0\ ,\quad
%g(k)=-ie^{i\theta_k}\tanh\Bigl(\frac{\beta}{2}\Bigr)
g(k)=-ie^{i\theta_k}\tanh\Bigl(\frac{\beta\varepsilon_h(k)}{2}\Bigr)\, . 
\ee
%%%%%%%%%%%%%%%%%%
\subsection{GGE Density Matrix} %
%%%%%%%%%%%%%%%%%%
It was shown in Ref.~[\onlinecite{CEF2:2012}] (see also
[\onlinecite{CEF:2011,CEF1:2012}]) that the RDM of the generalized Gibbs
ensemble \fr{rhoGGE},\fr{rhoGGE2} is Gaussian and can be expressed in
the form \eqref{eq:gaussian}, \fr{WGamma}. The correlation matrix is
given by
\be
\left(\Gamma_{\rm GGE}\right)_{ij}
= \frac{1}{{\cal Z}}\tr{}{e^{-\sum_{i,\sigma} \lambda^\sigma_j I^\sigma_j}\, a_{j}
    a_i}-\delta_{i j}\, .
\ee 
It can be written in the form \fr{eq:Gamma} with
\be
 \label{eq:gGGE}
f(k)=0\ ,\quad
g(k)=-ie^{i\theta_k}\tanh\Bigl(\frac{\gamma_k}{2}\Bigr)\, . 
\ee
Here the $\gamma_k$'s are related to the $\lambda_m^\sigma$'s by
\fr{gammak} and the Bogoliubov angle $\theta_k$ is given in \fr{theta_k}.

%%%%%%%%%%%%%%%%%%%%%%%
\subsection{Truncated GGE Density Matrix}%
%%%%%%%%%%%%%%%%%%%%%%%
The correlation matrix of the truncated GGE defined in
Sec.~\ref{sec:PGGE} is given by 
\be
\left(\Gamma_{\rm tGGE}^{(y)}\right)_{i j}=\frac{1}{\mathcal Z_y}\tr{}{\exp\Bigl(\sum_{n=0}^{y-1}\sum_{\sigma=\pm}[\lambda_{n,y}^\sigma I_n^\sigma]\Bigr)\, a_{j} a_i}-\delta_{i j}\, .
\ee
It can be written in the form \fr{eq:Gamma} with
\be
 \label{eq:tGGE}
f(k)=0\ ,\quad
g(k)=-ie^{i\theta_k}\mathrm{tanh}\Bigl(P_{y-1}(\cos(y))\varepsilon(k)\Bigr)\, .
\ee
Here $P_{y-1}(x)$ is a polynomial of order $y-1$, which is computed
numerically.

%%%%%%%%%%%%%%%%%%%%%%%
\subsection{Defective GGE Density Matrix}%
%%%%%%%%%%%%%%%%%%%%%%%
In Sec.~\ref{sec:DGGE} we defined the defective GGE $\rho_{\rm
  dGGE}^{(q)}$ as the ensemble that lacks in the conservation law
$I_q^{+}$. Its correlation matrix is given by
\be
\left[\bar \Gamma_{\rm dGGE}^{(+q)}\right]_{ij}=
\frac{1}{\mathcal Z^{(\rm d)}_q}
\tr{}{\exp\Bigl(\displaystyle\sum_{\ontop{n=0}{n\neq q}}^\infty
\lambda_{n, (+q)}^{+} I_n^+\Bigr)\, a_{j} a_i}
-\delta_{i j}\, . 
\ee
It can be written in the form \fr{eq:Gamma} with $f(k)=0$ and (\emph{cf.} Eq.~\eqref{eq:gGGE})
\be
 \label{eq:dGGE}
g(k)=-ie^{i\theta_k}\Bigl[\tanh\Bigl(\frac{\gamma_k}{2}\Bigr)-\kappa_q^+ \frac{\cos(q k)}{\varepsilon(k)}\Bigr]\, , 
\ee
where $\kappa_q^+$ is computed numerically maximizing the entanglement entropy, which selects $\lambda_{q, (+q)}^{+}=0$ whenever it is allowed.
We note that the Fourier transform of Eq.~\eqref{eq:dGGE}, which is required to compute the correlation matrix~\eqref{eq:Gamma_l}, can be easily expressed in terms of the GGE correlators;  for  $|\ell|<q$ we have
\begin{multline}\label{eq:defdeltag}
\int_{-\pi}^{\pi}\frac{\mathrm d k}{2\pi}e^{-i \ell
  k}g(k)=\int_{-\pi}^{\pi}\frac{\mathrm d k}{2\pi}e^{-i \ell k}g_{\rm GGE}(k)+\\
\frac{i\kappa_q}{4 J}\mathrm{sgn}(\log h)h^{\ell-1}e^{-|\log h|q}\, .
\end{multline}
Since $\kappa_q^+$ is a bounded function of $q$
(\emph{cf}. Eq.~\eqref{eq:bounds1}), at fixed $\ell$ the fermionic
correlators approach the GGE ones at least exponentially fast in $q$. 

%%%%%%%%%%%%%%%%%%%
\subsection{Quench Density Matrix}%
%%%%%%%%%%%%%%%%%%%
At zero temperature the ground state phase diagram of the TFIC
exhibits ferromagnetic ($h<1$) and paramagnetic ($h>1$) phases,
separated by a quantum critical point. In the ferromagnetic phase the
$\mathbb{Z}_2$ symmetry of the Hamiltonian is broken
spontaneously. As we will see, this symmetry breaking has important
effects on the time evolution of the density matrix.

%%%%%%%%%%%%%%%%%%%%%%%%%%%%%%%%%
\subsubsection{Quenches originating in the paramagnetic phase} %
%%%%%%%%%%%%%%%%%%%%%%%%%%%%%%%%%
Here at $t>0$ the full quench density matrix is 
\be
\rho(t)=|\Psi_t\rangle\ \langle\Psi_t|\ ,
\ee
where the state $|\Psi_t\rangle$ is given in \fr{Psit_para}. 
As a result of the squeezed-state form of $|\Psi_t\rangle$, Wick's
theorem applies to averages calculated with respect to $\rho(t)$, and
RDMs are Gaussians of the form \eqref{eq:gaussian}, \fr{WGamma}, with
correlation matrix 
\be\label{eq:Gammat}
\Gamma(t)={}_{\rm NS}\braket{0|e^{i H_{\rm NS}t}a_{j} a_{i} e^{-i
    H_{\rm NS} t}|0}_{\rm NS}-\delta_{i j}\, .
\ee
This is of the form \fr{eq:Gamma} with
\begin{eqnarray}\label{eq:fg}
g(k)&=&-ie^{i \theta_k}\Bigl[\cos\Delta_k-i\sin\Delta_k
  \cos(2\varepsilon_h(k) t)\Bigr],\nn 
f(k)&=&\sin\Delta_k \sin(2\varepsilon_h(k) t),
\end{eqnarray}
where $e^{i\theta_k}$ is given by \fr{theta_k}.

%%%%%%%%%%%%%%%%%%%%%%%%%%%%%%%%%%%%%%%%
\subsubsection{Quenches originating in the ferromagnetic phase}
%%%%%%%%%%%%%%%%%%%%%%%%%%%%%%%%%%%%%%%%
Given the initial state \fr{ferroGS}, the post-quench density matrix
of the full system is
\bea
\rho(t)&=&|\Psi_t\rangle\langle\Psi_t|\ ,\nn
\ket{\Psi_t}&=&\frac{e^{-i H_{\rm R} t}|0;h_0\rangle_{\rm R}+e^{-i
    H_{\rm NS}t}|0;h_0\rangle_{\rm NS}}{\sqrt{2}}.
\eea
Importantly, RDMs are no longer Gaussian in this case. We will discuss
how to cope with this complication in section \ref{s:SSB}. It is known
\cite{CEF2:2012} that in the stationary state RDMs are Gaussian with
a correlation matrix equal to the $t\to\infty$ limit of \fr{eq:Gammat}.

%%%%%%%%%%%%%%%%%%%%%%%%%%%%%%%
\section{Distances on the Space of RDMs}\label{s:distance} %
%%%%%%%%%%%%%%%%%%%%%%%%%%%%%%%
In the following we focus on RDMs for finite subsystems of lattice
models with a finite dimensional Hilbert space at each site. In this
case the RDMs are finite dimensional matrices and a simple way to
define a distance between two density matrices is by means of a matrix
norm
\be
d_a(\rho,\rho')=\parallel\rho-\rho'\parallel_a\, .
\ee
Here the index $a$ labels different matrix norms. As we are dealing
with finite matrices, all norms are equivalent in the sense that
\be\label{eq:equivalence}
c_{ab} \parallel\rho\parallel_a\leq \parallel\rho\parallel_b\leq
c_{ba}^{-1} \parallel\rho\parallel_a\, , 
\ee
where $c_{ab}$ and $c_{ba}$ are positive numbers that depend on the
matrix dimension but are independent of $\rho$. One consequence of
\fr{eq:equivalence} is that if the distance between two matrices
approaches zero when some external parameter $p$ is tuned to a value
$\bar p$, the dependence of the distance on $p-\bar p$ is almost
independent of the norm. On the other hand, the dependence on matrix
dimension is in general very different for different norms. This is
important for our purposes, because the matrix dimension is related to
the size of the subsystem under consideration, and it is principally
desirable to be able to compare distances between different sizes.

From a technical point of view, the distance induced by the  Frobenius
norm\footnote{It is also known as the ``Hilbert-Schmidt norm'', but we
prefer to call it ``Frobenius norm'' to emphasize that we are
considering finite subsystems.} 
\be
||A||_F\equiv \sqrt{\tr{}{A^\dag    A}}
\ee
is generally the easiest to calculate. On the other hand, it has the
drawback that the physical interpretation of the distance is less
transparent than for some other norms. For instance, given two density
matrices $\rho$ and $\rho'$, a very natural question is how different
expectation values of local observables are in the two ensembles. 
We now discuss this question for the particular case of spin-1/2
quantum spin chains. Here the most important local observables are 
products of Pauli matrices. These are particular cases of involutions
$\hat O^2=\mathrm I$, for which the following inequality holds
\be\label{eq:trnorm}
|\btr{}{(\rho-\rho') \hat O}|%\leq |\btr{}{|(\rho-\rho') \hat  O|}
\leq \parallel\rho-\rho'\parallel_1 \, . 
\ee 
Here 
\be
||A||_1\equiv \btr{}{\sqrt{A A^\dag}}
\ee
is the trace norm. From Eq.~\eqref{eq:trnorm} it is evident that the
trace distance provides an upper bound for the difference between the
expectation values of observables in the two states: if
$\parallel\rho-\rho'\parallel_1<\epsilon$, then the expectation
values of all (local) observables will agree in the two ensembles
within accuracy $\epsilon$.
In terms of the Frobenius distance we have instead (here we use that the
local Hilbert space is two-dimensional)
\be\label{eq:opF}
|\btr{}{(\rho-\rho') \hat O}|\leq
\parallel\rho-\rho'\parallel_1\leq 2^{\ell/2}
\parallel\rho-\rho'\parallel_F\, . 
\ee
On the other hand, we have
\be
|\btr{}{(\rho-\rho') \hat O}|\leq
|\btr{}{\rho\hat O}|+|\btr{}{\rho'\hat O}|\leq 2,
\label{bound=2}
\ee
where in the last step we have used that for involutions $\hat{O}$
\be
|\btr{}{\rho\hat O}|\leq\sum_j|\left(\rho\right)_{jj}|=
{\rm tr}\sqrt{\rho^\dagger\rho}=1.
\ee
Combining \fr{bound=2} and \fr{eq:opF} we see that 
as long as $ \parallel\rho-\rho'\parallel_F\gtrsim 2^{1-\ell/2}$,
the Frobenius distance does not provide useful information about
expectation values. It is shown in Appendix \ref{app:frob} that
for sufficiently large $\ell$ this is always the case.

A second problem with using the Frobenius norm as a distance is that
the norms of RDMs at late times after a quantum quench, as
well as the norms of RDMs describing Gibbs or generalized Gibbs
ensembles, generally are exponentially small in the subsystem
size. Given the upper bound derived in Appendix \ref{app:frob}
\be
\parallel\rho-\rho'\parallel_F
\leq\sqrt{\parallel\rho\parallel_F^2+\parallel\rho'\parallel_F^2},
\ee
this implies that in these cases of interest
$\parallel\rho-\rho'\parallel_F$ is exponentially small in subsystem
size. This shows that the Frobenius norm itself is not a convenient measure
for the distance between two density matrices. The same problem occurs for
the operator norm $||A||_{\rm op}=\sqrt{\lambda_{\rm max}}$, where
$\lambda_{\rm max}$ is the largest eigenvalue of
$A^\dagger A$. This norm was used for example in
Ref.~\cite{bhc-10} to analyze the relaxation properties of small
subsystems after a quench into a nonintegrable model. Indeed we have 
\be
\parallel \rho-\rho'\parallel_{op}\leq \parallel
\rho\parallel_{op}+\parallel\rho'\parallel_{op}\ ,
\ee
and the maximal eigenvalues of RDMs for large subsystems are generally
exponentially small in subsystem size.
%%%%%%%%%%%%%%%%%%%%
\subsection{Definition of the Distance}%
%%%%%%%%%%%%%%%%%%%%
In order to circumvent the problem described above, we define our
``distance''\footnote{We have not proven that ${\cal D}(\rho,\rho')$
obeys the triangle inequality as this is not essential for our
purposes.} 
on the space of RDMs as 
\be\label{eq:normD1}
\mathcal D(\rho,\rho')\equiv\frac{\parallel\rho-\rho'\parallel_F}
{\sqrt{\parallel \rho\parallel_F^2+\parallel\rho'\parallel_F^2}}\, .
\ee

\noindent\underline{1. An upper bound:}
Using the upper bound derived in \eqref{eq:upper_bound} we see that
\be
{\cal D}(\rho,\rho')\leq 1.
\ee 
\noindent\underline{2. A lower bound:}
A lower bound for $\mathcal D(\rho,\rho')$ can be established by means
of the triangle inequality $||\rho-\rho'||_F\geq |\ ||\rho||_F
-||\rho'||_F\ |$. Using that the Frobenius norm of a RDM is related to
the second R\'enyi entropy by
\be\label{eq:S2blah}
S_2\equiv -\log \tr{}{\rho^2}=-\log ||\rho||^2_F\ ,
\ee
we find that
\be
||\rho-\rho'||\geq
\left|\exp\left(-\frac{S_2}{2}\right)-\exp\left(-\frac{S'_2}{2}\right)\right|.
\ee
This provides the desired lower bound
\be
\mathcal D(\rho,\rho')\geq
\frac{|e^{-S_2/2}-e^{-S'_2/2}|}{\sqrt{e^{-S_2}+e^{-S'_2}}}\, .  
\label{lowerbound}
\ee
We note that the bound \fr{lowerbound} is independent of subsystem
size $\ell$ as long as the second R\'enyi entropies of the two
ensembles differ at least by a constant (when viewed as functions of
$\ell$). 

%%%%%%%%%%%%%%%%%%%%%%%%%%%%%%%%%%%%
\subsection{The Distance between two Thermal Ensembles}\label{s:dt}%
%%%%%%%%%%%%%%%%%%%%%%%%%%%%%%%%%%%%
In order to establish a benchmark for \fr{eq:normD1},
it is useful to consider the distance between the RDMs of two thermal
ensembles at slightly different inverse temperatures $\beta$ and
$\beta'$ (but the same Hamiltonian). Then
\be
{\cal
  D}(\rho_{\beta},\rho_{\beta'})\approx
\frac{||\frac{\partial\rho_\beta}
{\partial\beta}||_F}{||\rho_\beta||_F}\ \frac{1}{\sqrt{2}}|\beta-\beta'|.
\ee
For a sufficiently large subsystem (and a local Hamiltonian), the
first factor on the right hand side can be expressed as
\bea
\frac{||\frac{\partial\rho_\beta}{\partial\beta}||_F^2}{||\rho_\beta||_F^2}&=&
\frac{||\rho_\beta\left(\langle H\rangle_\beta-H\right)||_F^2}
{||\rho_\beta||_F^2}\nn
&=&\frac{{\rm Tr}\left[\rho_\beta^2\left(\langle H\rangle_\beta-H\right)^2\right]}
{{\rm Tr}\rho_\beta^2}\nn
&=&\big\langle\left(\langle H\rangle_\beta-H\right)^2\big\rangle_{2\beta},
\eea
where $\langle{\cal O}\rangle_\beta={\rm Tr}\big(\rho_\beta{\cal
  O}\big)$. For a large subsystem, this is proportional to the square
of its size, and hence
\be
\frac{||\frac{\partial\rho_\beta}{\partial\beta}||_F}{||\rho_\beta||_F}
\propto\ell.
\ee
We conclude that the distance between two thermal RDMs on a subsystem
of size $\ell$, and $\beta\approx\beta'$ is
\be
{\cal D}(\rho_{\beta},\rho_{\beta'})\propto
\ell|\beta-\beta'|.
\ee
As expected this is proportional to the difference in temperatures,
but there is also a factor of $\ell$. The latter is important if one
is interested in comparing the distance between two ensembles for
different subsystem sizes.
%%%%%%%%%%%%%%%%%%%%%%%%%%%%%%%%
\subsection{The Distance between two GGEs}\label{s:dGGE}%
%%%%%%%%%%%%%%%%%%%%%%%%%%%%%%%%
The above discussion carries over to the case of two generailzed Gibbs
ensembles \fr{rhoGGE}, with slightly different values of Lagrange multipliers
$\lambda_m^\sigma$. The leading contribution to the distance is given by
\be\label{eq:dGGEscal}
{\cal D}(\rho_{\rm GGE},\rho_{\rm GGE}')\approx
\sum_{m,\sigma}
\frac{||\frac{\partial\rho_{\rm GGE}}
{\partial\lambda_m^\sigma}||_F}{||\rho_{\rm GGE}||_F}\ 
\frac{1}{\sqrt{2}}|\lambda_m^\sigma-{\lambda'}_m^\sigma|.
\ee
A calculation similar to the thermal case shows that for large
subsystem size
\bea
\frac{||\frac{\partial\rho_{\rm
      GGE}}{\partial\lambda_m^\sigma}||_F}{||\rho_{\rm GGE}||_F}\propto\ell.
\eea

%%%%%%%%%%%%%%%%%%%%%%%%%%%%%%%%
\subsection{Information on observables contained in the distance}
\label{s:Dobs}%
%%%%%%%%%%%%%%%%%%%%%%%%%%%%%%%%
Let us consider the situation where the distance between two reduced
density matrices $\rho_1$ and $\rho_2$ defined on an interval of
length $\ell$ becomes small, and denote the corresponding averages of
local operators on said interval by
\be
\langle {\cal O}\rangle_a={\rm Tr}\left[\rho_a{\cal O}\right] ,\quad a=1,2.
\ee
By expanding the density matrices in a complete basis of Hermitian
involutions we can show that
\be\label{eq:relDnorm}
\mathcal D(\rho_{1},\rho_{2})= 
\sqrt{\frac{\sum_{\mathcal O}(\braket{\mathcal O}_2-\braket{\mathcal
      O}_1)^2}{\sum_{\mathcal O}(\braket{\mathcal
      O}_2^2+\braket{\mathcal O}_{1}^2)}}.
\ee
Defining an average
\bea
\overline{f({\cal O})}&\equiv&\sum_{\cal O}P({\cal O}) f({\cal
  O})\ ,\nn
P({\cal O})&=&\frac{\braket{\mathcal O}^2_1+\braket{\mathcal O}^2_{2}}
{\sum_{\cal Q }\braket{\mathcal Q}^2_1+\braket{\mathcal Q}^2_{2}},
\label{PofO}
\eea
we can express the distance \fr{eq:normD1} as
\be
\label{eq:Dsur}
\mathcal D(\rho_1,\rho_2)=
\left({\overline{\left(R({\mathcal O})\right)^2}}\right)^{1/2}.
\ee
Here
\be
R({\mathcal O})\equiv 
\frac{|\braket{\mathcal O}_1-\braket{\mathcal O}_2|}
{\sqrt{\braket{\mathcal O}_1^2+\braket{\mathcal O}_2^2}}.
\ee
is the relative difference between the ensembles described by $\rho_1$
and $\rho_2$, respectively. This implies that ${\cal
  D}(\rho_1,\rho_2)$ measures the mean relative difference of the
expectation values of all local operators, averaged with respect to
the probability distribution \fr{PofO}.

%%%%%%%%%%%%%%%%%%%%%%%%%%%%%%%%%
\subsection{The Distance between two Gaussian Density Matrices}  %
%%%%%%%%%%%%%%%%%%%%%%%%%%%%%%%%%
The distance \fr{eq:normD1} between two Gaussian RDMs $\rho[\Gamma]$ and
$\rho[\Gamma']$ can be expressed in terms of their correlation matrices
following Ref.~[\onlinecite{FC:2010}]. Given the definition of the
distance 
\bea
{\cal  D}\big(\rho,\rho'\big)&=&
\frac{\sqrt{{\rm Tr}\big(\rho^2+{\rho'}^2-2\rho\rho'\big)}}
{\sqrt{{\rm Tr}\big(\rho^2\big)+{\rm Tr}\big({\rho'}^2\big)}},
\label{dist}
\eea
we require tractable expressions for the quantities
\be
{\rm Tr}\big(\rho[\Gamma]\rho[\Gamma']\big).
\ee
This is achieved in two steps. First, we note that the product of two
Gaussian RDMs \fr{eq:gaussian} is itself Gaussian\cite{FC:2010}
%\begin{widetext}
\bea
&&\exp\left(\frac{1}{4}\sum_{i,j}W_{i j}a_i a_j\right)
\exp\left(\frac{1}{4}\sum_{i,j}\widetilde{W}_{i j}a_i a_j\right)\nn
&&=\exp\left(\frac{1}{4}\sum_{i,j}[\log(e^We^{\widetilde{W}})]_{i j}a_i a_j\right).
\label{prodG}
\eea
%\end{widetext}
This can be seen by expanding the left hand side of \fr{prodG} by
means of the Baker-Campbell-Hausdorff formula in terms of multiple
commutators, and then observing that the commutator of quadratic
operators is quadratic and gives rise to a commutator between the
matrices $W$ that characterize the 2-forms
\be\label{eq:commquad}
[\frac{1}{4}\sum_{i,j}W_{i j}a_i a_{j},\frac{1}{4}\sum_{i,j}\widetilde W_{i j}a_i a_{j}]=\frac{1}{4}\sum_{i,j}\Bigl[[W,\widetilde W]\Bigr]_{i j} a_i a_j\, .
\ee
Second, using results for the second R\'enyi entropy
\cite{Renyi2}, one can relate the Frobenius norm of a Gaussian RDM to
the correlation matrix by 
\be
{\rm Tr}\big[\rho[\Gamma]^2\big]=\Bigl({\rm det}\Bigl|\frac{\mathrm I+\Gamma^2}{2}\Bigr|\Bigr)^\frac{1}{2}.
\label{trrho2}
\ee
Combining \fr{trrho2} and \fr{prodG} one can then show~\cite{FC:2010} that
\be\label{eq:trbasic}
\{\Gamma,\tilde\Gamma\}\equiv\btr{}{\rho[\Gamma]\rho[\tilde \Gamma]}
=\Bigl(\det\Bigl|\frac{\mathrm I+\Gamma\tilde\Gamma}{2}\Bigr|\Bigr)^\frac{1}{2}\, .
\ee
Here we have used that $\tr{}{\rho_1\rho_2}\geq 0$, which is a
consequence of density matrices being positive semidefinite
operators. Finally, substituting \fr{eq:trbasic} into \fr{dist}, we
obtain the following result for the distance between two Gaussian RDMs
\be\label{eq:DFfree}
\mathcal
D(\rho[\Gamma],\rho[\tilde\Gamma])=\left[1-\frac{2\{\Gamma,\tilde\Gamma\}}{\{\Gamma,\Gamma\}+\{\tilde\Gamma,\tilde\Gamma\}}\right]^\frac{1}{2}. 
\ee
Given that the correlation matrices are only $2\ell$ dimensional
(with $\ell$ the subsystem size), \eqref{eq:DFfree} provides a very
efficient way of computing distances for large subsystem sizes.

%%%%%%%%%%%%%%%%%%%%%%%%%
\section{Single-site subsystem}\label{sec:1spind}%
%%%%%%%%%%%%%%%%%%%%%%%%%
It is instructive to consider the time evolution of the RDM describing
a single-site subsystem in some detail. In this case the RDM of site
$1$ can be
expressed in the form
\be
\rho_1(t)=\frac{\mathrm I}{2}+\vec m(t)\cdot\vec \sigma_1\, ,
\ee
where $\vec m(t)$ is the magnetization per site at time $t$ after the
quench, i.e.
\be
m^\alpha(t)=\frac{1}{2}\langle\Psi_t|\sigma_1^\alpha|\Psi_t\rangle.
\ee
The RDM of the generalized Gibbs ensemble describing the stationary
state is
\be
\rho_{\rm GGE,1}=\frac{\mathrm I}{2}+m^z_{\rm stat}\sigma^z_1\, .
\ee
where
\be
m^z_{\rm stat}=\int_{-\pi}^\pi\frac{dk}{4\pi}
  e^{i\theta_k}\cos\Delta_k.
\ee
Finally, the RDM of the thermal ensemble described by $\rho_\beta$,
whose inverse temperature $\beta$ is fixed by the requirement
\be
\lim_{L\to\infty}\frac{1}{L}
\langle \Psi_0|H(h)|\Psi_0\rangle=
\lim_{L\to\infty}\frac{1}{L}{\rm Tr}\left[\rho_\beta H(h)\right],
\ee
is given by
\be
\rho_{\beta,1}=\frac{\mathrm I}{2}+m^z_\beta\sigma^z_1\, .
\ee
Here the transverse magnetization per site is
\be
m^z_\beta=\int_{-\pi}^\pi\frac{dk}{4\pi}
  e^{i\theta_k}\tanh\big(\frac{\beta\varepsilon_k}{2}\big).
\ee
%%%%%%%%%%%%%%%%%%%%%%%%%%%%%%%
\subsection{Quenches originating in the paramagnetic phase}%
%%%%%%%%%%%%%%%%%%%%%%%%%%%%%%%
Here the $\mathbb{Z}_2$ symmetry enforces
\be
m^x(t)=m^y(t)=0.
\ee
The z-component of the magnetization per site is
\be
m^z(t)=\int_{-\pi}^\pi\frac{dk}{4\pi}
  e^{i\theta_k}\left[\cos\Delta_k-i\sin\Delta_k\cos(2\varepsilon_k t)\right].
\label{mz}
\ee
For late times we may evaluate the integral by means of a stationary
phase approximation, which gives
\be\label{eq:mz1}
m^z(t)\simeq m^z_{stat}+\frac{c(t)}{(Jt)^{3/2}}\ ,
\ee
where
\bea
c(t)&=&\frac{(h-h_0)\cos(4Jt|1-h|-\pi/4)}{8|h_0-1|\sqrt{\pi|h-1|}}\nn
&+&\frac{(h-h_0)\cos(4Jt|1+h|+\pi/4)}{8|h_0+1|\sqrt{\pi|h+1|}}.
\eea
The distance between $\rho_1(t)$ and the generalized Gibbs RDM at
late times then decays to zero like a power-law with exponent $3/2$
\bea
{\cal D}(\rho_1(t),\rho_{\rm
  GGE,1})&=&\frac{\sqrt{2}|m^z(t)-m^z_{stat}|}
{\sqrt{1+2(m^z(t))^2+2(m^z_{\rm GGE})^2}}\nn
&\sim&\sqrt{\frac{2c^2(t)}{1+4(m^z_{stat})^2}}(Jt)^{-\frac{3}{2}}.
\label{rho1_GGE}
\eea
On the other hand, the distance between $\rho_1(t)$ and the thermal
RDM approaches a constant at late times
\begin{multline}
{\cal D}(\rho_1(t),\rho_{\beta,1})=\frac{\sqrt{2}|m^z(t)-m^z_\beta|}{\sqrt{1+2(m^z(t))^2+2(m^z_\beta)^2}}\\
\sim\frac{\sqrt{2}|m^z_{stat}-m^z_\beta|}{\sqrt{1+2(m^z_{stat})^2+2(m^z_{\beta})^2}}+{\cal
  O}\big((Jt)^{-\frac{3}{2}}\big)\, .
\label{rho1_thermal}
\end{multline}
%%%%%%%%%%%%%%%%%%%%%%%%%%%%%%%%%%%%%%%
\subsection{Quenches originating in the ferromagnetic phase}\label{s:ferro1} %
%%%%%%%%%%%%%%%%%%%%%%%%%%%%%%%%%%%%%%%
Here all three components of the magnetization per site are non-zero.
The component along the transverse field direction is again given by
\fr{mz}, while the late-time asymptotics of $m^x(t)$ has been
calculated in \cite{CEF1:2012}
\be\label{eq:mx}
m^x(t)=\frac{1}{2}\sqrt{\mathcal C^x_{FF}}e^{t \int_0^\pi\frac{\mathrm
    d k}{\pi}\log\cos\Delta_k\varepsilon^\prime_k}\, . 
\ee
Here $\mathcal C^x_{FF}$ is a known amplitude and
$\varepsilon^\prime_k=\frac{d\varepsilon_h(k)}{dk}$. Finally, the
Heisenberg equation of motion for $\sigma^x_1(t)$ relates the $y$ and
$x$ components  
\be
\label{eq:my}
m^y(t)=\frac{1}{2 J h}\frac{dm^x(t)}{dt}.
\ee
Importantly, $m^{x,y}(t)$ exhibit exponential decay in time. In
contrast, $m^z(t)$ again decays like a power law with exponent $3/2$
and therefore will dominate the late time behaviour. Hence at
sufficiently late times, the distances of $\rho_1(t)$ to GGE and
thermal RDMs are again given by \fr{rho1_thermal} and \fr{rho1_GGE}
respectively. So for a single site subsystem the spontaneous symmetry
breaking only modifies the intermediate time behaviour of the
distances. As we will see, this holds true also for larger subsystems.

%%%%%%%%%%%%%%%%%%%%%%%%%%%%%%%%%%%%%
\section{Larger Subsystems for Quenches from the Paramagnetic Phase}  %
\label{sec:larger}
%%%%%%%%%%%%%%%%%%%%%%%%%%%%%%%%%%%%%

\begin{figure}[ht]
\begin{center}
\includegraphics[width=0.45\textwidth]{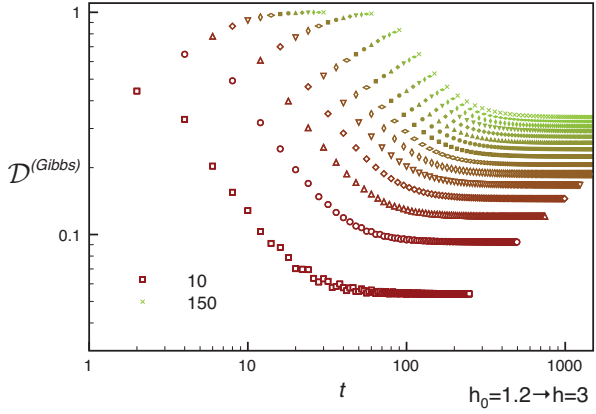}
\caption{
Normalized distance ${\mathcal D}^{\rm Gibbs}=
\mathcal D(\rho_\ell(t),\rho^{\beta}_\ell)$ 
after a quench within the paramagnetic phase for subsystem sizes
$\ell=10,20,\dots,150$. As $\ell$ increases, the color fades from
brown to green, the symbols become smaller and the curves narrower. 
At late times the distances tend to constants depending on subsystem
size. 
}\label{fig:para_gibbs}
\end{center}
\end{figure}

\begin{figure}[ht]
\begin{center}
\includegraphics[width=0.45\textwidth]{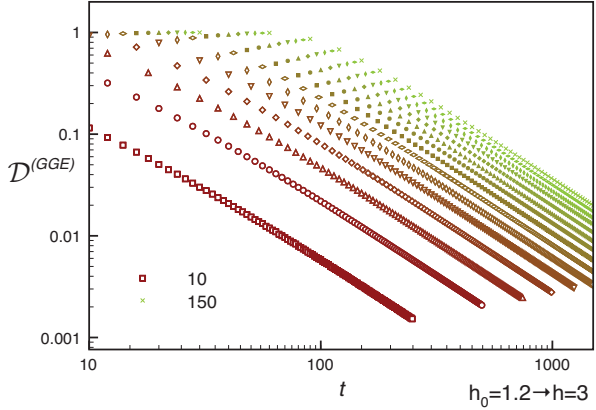}
\caption{
Normalized distance ${\mathcal D}^{\rm GGE}=\mathcal
D(\rho_\ell(t),\rho^{GGE}_\ell)$ 
after a quench within the paramagnetic phase for subsystem sizes
$\ell=10,20,\dots,150$. As $\ell$ increases, the color fades from
brown to green, the symbols become smaller and the curves narrower. At
late times $D(\rho_\ell(t),\rho^{GGE}_\ell)$ tends to zero 
in a universal power-law fashion $\propto (Jt)^{-3/2}$.
}\label{fig:para_gge}
\end{center}
\end{figure}

\begin{figure}
\begin{center}
\includegraphics[width=0.45\textwidth]{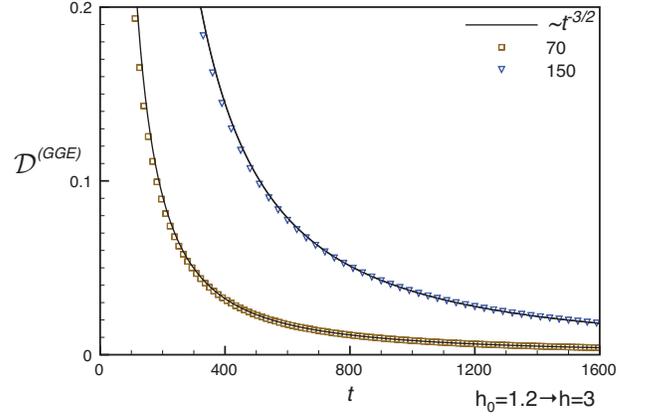}
\caption{Distance 
${\mathcal D}^{\rm GGE}=\mathcal D(\rho_\ell(t),\rho^{GGE}_\ell)$
after a quench within the paramagnetic phase
for two representative values $\ell=70,150$.  
We used the same notations of Fig.~\ref{fig:para_gge}. The black
dashed curves are best fits to the form $\mathcal D=a t^{-3/2}$.
}\label{fig:DGGEpower}
\end{center}
\end{figure}

\begin{figure}
\begin{center}
\includegraphics[width=0.45\textwidth]{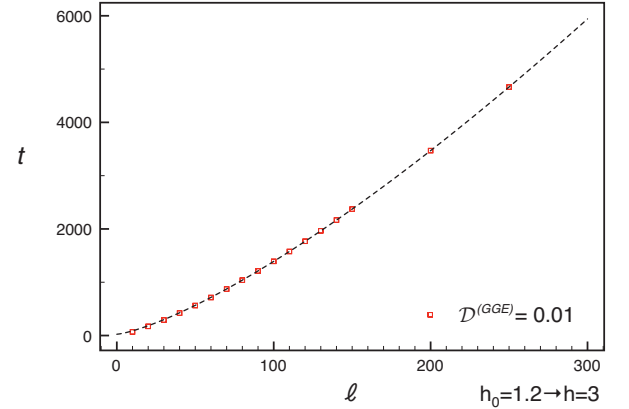}
\caption{Dependence of time on subsystem size at fixed distance
$\mathcal D(\rho_\ell(t),\rho^{GGE}_\ell)=0.01$ for the same
parameters as in Fig.~\ref{fig:para_gge}. The dashed curve is the best
fit to the functional form $t=a+b\ell^{4/3}$.
}\label{fig:chipara}
\end{center}
\end{figure}

For quenches with $h_0>1$ and in the thermodynamic limit, we determine
the distance between the quench RDM and that of an appropriate thermal
or generalized Gibbs ensemble by means of relation \fr{eq:DFfree}. The
correlation matrices for all cases are of the form \fr{eq:Gamma},
\fr{eq:Gamma_l} with elements given in \fr{eq:thermal}, \fr{eq:gGGE}
and \fr{eq:fg} respectively. For a subsystem of size $\ell$ this
requires the calculation of determinants of $2\ell\times 2\ell$
matrices, which is done numerically. Results for a quench from
$h_0=1.2$ to $h=3$ and subsystem sizes $\ell=10,20,30\ldots,150$ are 
shown in Figs~\ref{fig:para_gibbs} and \ref{fig:para_gge}
We see that the distance between quench and Gibbs RDMs tends to a
$\ell$-dependent constant at late times. This establishes that
subsystems do not thermalize. On the other hand, as can be seen from
Fig.~\ref{fig:para_gge}, at sufficiently late times the distance
between $\rho_\ell(t)$ and $\rho_{\rm GGE,\ell}$ decays to zero in a
universal power-law fashion
\be\label{eq:Dnum}
\mathcal D(\rho_\ell(t),\rho_{\rm GGE,\ell})\xrightarrow{J t\gg
  1}k(\ell)(J t)^{-3/2}+\dots\, .
\ee
The quality of the fit \fr{eq:Dnum} is shown in Fig.~\ref{fig:DGGEpower}.
The large-$\ell$ asymptotics of the function $k(\ell)$ can be inferred
as follows. On surfaces with constant, small $\mathcal D$ the time scales
as $t\sim \ell^{4/3}$ as is shown in Fig.~\ref{fig:chipara}. This in
turn implies that
\be\label{eq:Dnum1}
k(\ell)\sim \ell^{2}\, .
\ee

%%%%%%%%%%%%%%%
\subsection{Relaxation time}%
%%%%%%%%%%%%%%%
We may extract a relaxation time from the behaviour of the distance, by
using the connection to averaged differences in the expectation values
of local operators established in subsection \ref{s:Dobs}. The
distance can be written as
\be
\mathcal D(\rho_{{\rm GGE},\ell},\rho_\ell(t))= 
\left(\overline{\left[R({\cal O})\right]^2}\right)^{1/2},
\ee
where
\be
R({\cal O})\equiv \frac{|\braket{\mathcal O}_t-\braket{\mathcal
    O}_{\rm GGE}|}{\sqrt{\braket{\mathcal O}_t^2+\braket{\mathcal
      O}_{\rm GGE}^2}},
\ee
and the bar denotes the average \fr{PofO}. Using that
\be
\label{eq:rmst}
\overline{R({\cal O})}\leq\sqrt{\overline{
\left[R({\cal O})\right]^2}}
=\mathcal D(\rho_{{\rm GGE},\ell},\rho_\ell(t)),
\ee
and then substituting the asymptotic behaviour \fr{eq:Dnum},
\fr{eq:Dnum1} into the right hand side, we obtain
\be
\overline{R({\cal O})}\alt \ell^2t^{-3/2}\ .
\ee
Bounding the right hand side by a (small) constant, we obtain a time
scale $t^*_{rms}$ associated with the relaxation of the average
relative error with respect to the distribution \eqref{PofO}
\be\label{eq:relaxrms}
t^{\ast}_{\rm rms}\sim\ell^{4/3}\, .
\ee
It is not simple to identify the observables that give significant
contribution to the average, since it depends both on their
``multiplicity''  in the subsystem (produced by translational
invariance and other symmetries) and on the expectation values. 
We note that the relaxation time $t^*_{\rm rms}$ is very different
from the time scales identified in Ref.~[\onlinecite{CEF2:2012}] in
the time evolution of the two point functions of spin operators for
quenches within the paramagnetic phase.

%%%%%%%%%%%%%%%%%%%%%%%%%%%%%%%%%%%%%%%%%%%%%%%%%%%%
\subsection{Distance from Truncated Generalized Gibbs Ensembles}
\label{s:GGEy}
\begin{figure}[ht]
\begin{center}
%\vspace{0.25cm}
\includegraphics[width=0.45\textwidth]{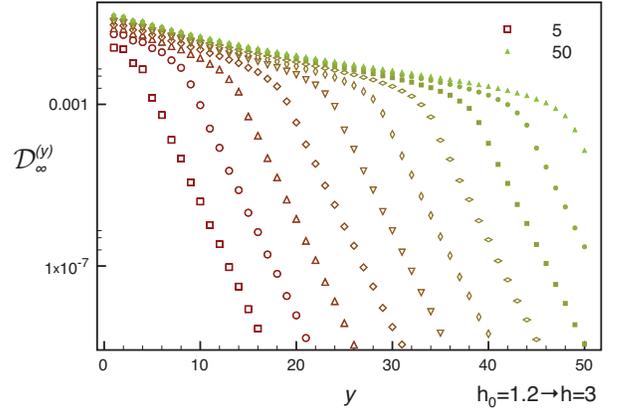}
\caption{Distance $\mathcal D_\infty^{(y)}=\mathcal
  D(\rho_{GGE,\ell},\rho^{(y)}_{tGGE,\ell})$ between the GGE and the
truncated GGEs obtained by imposing local conservation laws with
densities involving at most $y+1$ consecutive sites. The quench is
from $h_0=1.2$ to $h=3$ and the subsystem size ranges from
$\ell=5$ to $\ell=50$. Colors and sizes change gradually as a function
of the size $\ell$. For $y>\ell$, the distance starts decaying
exponentially in $y$, with an $\ell$-independent decay constant.
}\label{fig:local}
\end{center}
\end{figure}

%%%%%%%%%%%%%%%%%%%%%%%%%%%%%%%%%%%%%%%%%%%%%%%%%%%%%
Having established that the distance between quench and GGE reduced
density matrices tends to zero as a universal power law at late times,
a natural question is, how close the quench RDM is to the truncated GGEs
\fr{rho_tGGE}, which retain only finite numbers of conservation laws.
\begin{figure}[ht]
\begin{center}
%\vspace{0.25cm}
\includegraphics[width=0.45\textwidth]{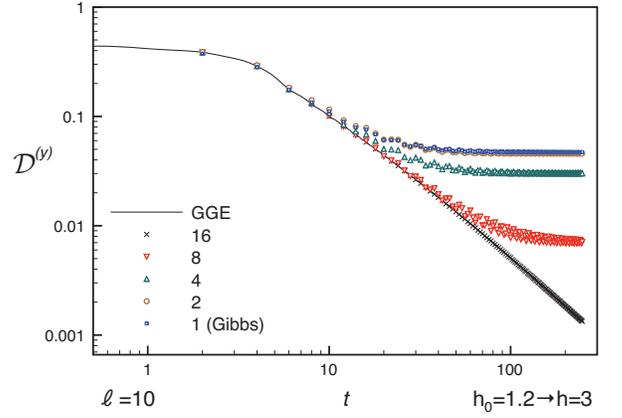}
\caption{Distance $\mathcal D^{(y)}=\mathcal D(\rho_\ell(t),\rho_{{\rm
tGGE},\ell}^{(y)})$  at fixed length $\ell=10$ between quench and
truncated GGE reduced density matrices for $y=1,2,4,8,16$ and a
quench within the paramagnetic phase. Here $y$ is the maximal range
of the densities of local conservation laws included in the definition
of the ensemble. As the number of conservation laws is increased, the
time window, in which the distance decays as $t^{-3/2}$, increases.
At very late times all distances with finite $y$ saturate to nonzero
values.
}\label{fig:cons}
\end{center}
\end{figure}
A representative example for a quench within the paramagnetic phase is
shown in Fig.~\ref{fig:cons}. We see that at sufficiently late times,
the distances converge to constant values. However, increasing the
range (and number) of conservation laws, the values of these plateaux
decrease, signalling that retaining more conservation laws gives
better descriptions. In an intermediate time window, the extent of
which grows with $y$, the distance decays in a universal $t^{-3/2}$
power-law fashion. 
In Fig.~\ref{fig:local} we consider the distance 
\be\label{eq:truncfig}
{\cal D}^{(y)}_\infty=
\lim_{t\to\infty}\mathcal D(\rho_\ell(t),\rho_{{\rm
    tGGE},\ell}^{(y)})
=\mathcal D(\rho_{{\rm GGE},\ell},\rho_{{\rm tGGE},\ell}^{(y)}),
\ee
between the RDMs of the truncated and full generalized Gibbs ensembles 
as a function of the parameter $y$. For a given subsystem size $\ell$,
this corresponds to plotting the  values of the plateaux seen in
Fig.~\ref{fig:cons} against the corresponding values of $y$.
The distance is seen to start decaying exponentially as a function of
$y$ as soon as $y\agt\ell$. 

There are two main conclusions of the above analysis:
\begin{enumerate}
\item{} Including more local conservation laws improves the description of
the stationary state.
\item{} The description of the stationary state improves rapidly, once
the range $y+1$ of all conservation laws not included in the truncated
GGE exceeds the subsystem size $\ell$.
\end{enumerate}

%%%%%%%%%%%%%%%%%%%%%%%%%%%%%%%%%%%%%%%%%%%%%%%%%%%%
\subsection{Distance from defective Generalized Gibbs Ensembles}
\label{s:GGEyq}
\begin{figure}[ht]
\begin{center}
\includegraphics[width=0.45\textwidth]{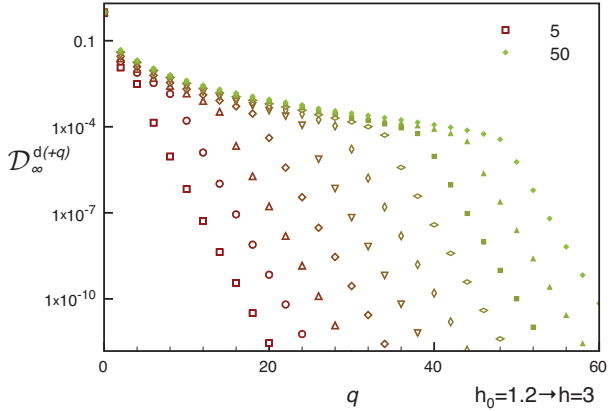}
\caption{Distance $\mathcal D_\infty^{{\rm d}(+q)}=\mathcal
D(\rho_{GGE,\ell},\rho^{(+q)}_{dGGE,\ell})$ for a quench within the
paramagnetic phase, for subsystem lengths $\ell=5,10,\dots,50$. The
excluded conservation law is $I^+_q$ with even $q$. Colors and sizes
change gradually as a function of the length. 
When $y>\ell$, the distance starts decaying exponentially with a decay
length given by Eq.~\eqref{eq:dGGEasympt}.  
}\label{fig:local1}
\end{center}
\end{figure}

%%%%%%%%%%%%%%%%%%%%%%%%%%%%%%%%%%%%%%%%%%%%%%%%%%%%%
We now turn to the role played by particular local conservation
laws. We find that the distance between quench and defective GGE
reduced density matrices for a given quench and subsystem size tends
to a constant at late times, i.e. 
\be
\lim_{t\to\infty}\mathcal D(\rho_\ell(t),\rho_{{\rm
    dGGE},\ell}^{(q)})\equiv{\cal D}_\infty^{{\rm d}(+q)}.
\ee
The dependence of this asymptotic value on the subsystem size $\ell$
and the integer $q$ is shown in Fig.~\ref{fig:local1} for a quench
within the paramagnetic phase.
We see that ${\cal D}_\infty^{{\rm d}(+q)}$ exhibits an exponential
decay in $q$ as soon as $q\agt\ell$. This is similar to the behaviour
observed in the truncated GGE case. The decay length can be calculated
from the large-$q$ asymptotics of Eq.~\eqref{eq:defdeltag}. By series
expanding Eq.~\eqref{eq:DFfree} to second order in $\Gamma_{\rm
  dGGE}^{(+q)}-\Gamma_{\rm GGE}$ we obtain 
\be
\label{eq:dGGEasympt}
{\cal D}_\infty^{{\rm d}(+q)}
\overset{y\gg\ell}{\sim} |\kappa_q^+| e^{-|\log h|(q-\ell)}\, .
\ee
Numerically we find that $\kappa_q^+\sim 1/q^2$.  

%%%%%%%%%%%%%%%%%%%%%%%%%%%%%%%%%%%%%%%%%%%%%%%%%%%%
\subsubsection{``GGE Reconstruction''}
\begin{figure}[ht]
\begin{center}
\includegraphics[width=0.45\textwidth]{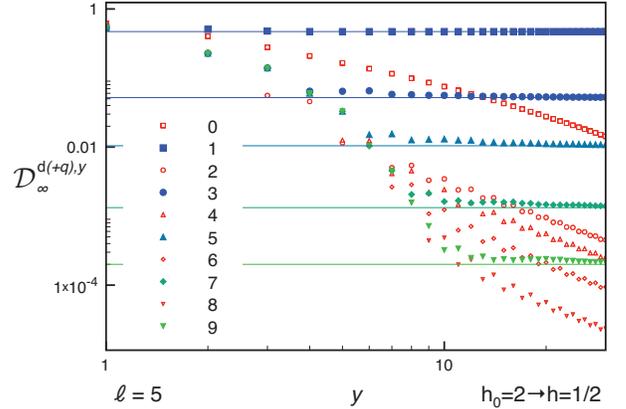}
\caption{
Distance $\mathcal D_\infty^{{\rm d}(+q),y}=\mathcal
D(\rho_{GGE,\ell},\rho^{(+q),y}_{dtGGE,\ell})$ for a quench across the
critical point between the GGE and the defective truncated GGE RDMs
for a subsystem of $5$ consecutive spins, as a function of the number
$y$ of retained conservation laws. Each symbol corresponds to a
different excluded conservation law $I^+_q$ (the legend indicates the
value of $q$). The distance approaches zero for even $q$, 
whereas it remains finite for odd $q$, in agreement with the discussion
of section~\ref{sec:DGGE}. The lines are the distances from the
corresponding defective generalized Gibbs ensemble $\rho_{{\rm
    dGGE}}^{(+q)}$ with maximal entanglement entropy. 
}\label{fig:local2}
\end{center}
\end{figure}

%%%%%%%%%%%%%%%%%%%%%%%%%%%%%%%%%%%%%%%%%%%%%%%%%%%%
In section~\ref{sec:DGGE} we discussed the issue that, for certain
quenches and omitted conservation laws $I_q^+$, the corresponding
defective GGE is identical to the full generalized Gibbs ensemble.
We now return to this point. In Fig.~\ref{fig:local2} we consider 
the truncated, defective GGE for a quench across the critical point
from $h_0=2$ to $h=0.5$ for a subsystem of length $\ell=5$. We plot
the distance between the reduced density matrices of the appropriate
GGE and the truncated, defective GGE with $y$ integrals of motion,
where $I_q^+$ ($q<y$) has been excluded, i.e.
\be
\mathcal D_\infty^{{\rm d}(+q),y}=\mathcal
D(\rho_{\rm GGE,\ell},\rho^{(+q),y}_{\rm tdGGE,\ell}).
\ee
As discussed in section~\ref{sec:DGGE}, for even $q$ we expect this
distance to approach zero, when the number $y$ of conservation laws
goes to infinity. This behaviour is clearly observed in
Fig.~\ref{fig:local2}. This implies that the corresponding
conservation laws do not affect averages of local operators. As
discussed before, this is a particular feature of free theories, where
$H(h_0)$ and $H(h)$ generically share certain local conservation laws.

%%%%%%%%%%%%%%%%%%%%%%%%%%%%%%%%%%%%%%%%%%%%%%%%%%%%%%%%%%%%%%%
\subsubsection{More local conservation laws are more important}
%%%%%%%%%%%%%%%%%%%%%%%%%%%%%%%%%%%%%%%%%%%%%%%%%%%%%%%%%%%%%%%

On the other hand, for odd $q$ we find that $\mathcal D_\infty^{{\rm
    d}(+q),y}$ approaches constant values when $y$ becomes large. This
value agrees with the distance between the GGE and the
defective GGE with maximal entanglement entropy (we stress that for
the considered quench the defective GGE does not always correspond to a
stationary point of the entanglement entropy under a variation of the
excluded integral of motion, as shown in Fig.~\ref{fig:defentropy} of Appendix~\ref{app:defective}).   

The fact that $\mathcal D_\infty^{{\rm  d}(+q),y}$ tends to a constant
at large $y$ shows that retaining an infinite number of local
conservation laws while excluding one of them is insufficient for
describing the stationary state. By comparing distances for different
values of $q$ we observe that for a given value of $y$, $\mathcal
D_\infty^{{\rm  d}(+q),y}$ decreases as a function of $q$. This implies
\emph{the more local the conservation
  law, the more important it is for describing the stationary state}.

%%%%%%%%%%%%%%%%%%%%%%%%%%%%%%%%%%%%%%%%%%%%%%%%%%%%%%%%%%%%%
\section{Quenches from the ferromagnetic phase: effects of
spontaneous symmetry breaking.
\label{s:SSB}}
%%%%%%%%%%%%%%%%%%%%%%%%%%%%%%%%%%%%%%%%%%%%%%%%%%%%%%%%%%%%%%%%%%%
We now turn to quenches originating in the ferromagnetic phase,
i.e. $h_0<1$. In this case, the time evolved initial state is given by
\bea
\ket{\Psi_t}&=&\frac{\ket{\psi_t}_{\rm R}+\ket{\psi_t}_{\rm
    NS}}{\sqrt{2}}\ ,\nn
\ket{\psi_t}_{a}&=&e^{-i H_{a} t}\ket{0;h_0}_{a}\quad a=\rm{R}, \rm{NS}\, ,
\eea
where $\ket{0;h_0}_{\rm R/NS}$ are the ground states of the
Hamiltonian $H(h_0)$ with periodic/antiperiodic boundary conditions.
In order to analyze reduced density matrices after a quantum quench
from the ferromagnetic phase we will make use of the following facts.
\begin{enumerate}[a)]
\item The fermion parity $e^{i\pi \mathcal N}=\prod_j\sigma_j^z$ Eq.~\eqref{eq:calN}  is fully factorized in space.
\label{e:a}
\item The states $\ket{\psi_t}_{\rm R}$  and $\ket{\psi_t}_{\rm NS}$ 
are eigenstates of $e^{i\pi \mathcal N}$ with eigenvalues $1$ and $-1$
respectively. 
\label{e:b}
\item The difference between the expectation values of local operators
in the states $\ket{\psi_t}_{\rm R}$ and $\ket{\psi_t}_{\rm NS}$ tends to
zero in the thermodynamic limit.
\label{e:c}
\item The RDMs ${\rm Tr}_{\bar{A}}\left[ |\psi_t\rangle_{\tt a}{}_{\tt
  a}\langle\psi_t|\right]$, ${\tt a}={\rm R,NS}$, where $A$ is a
single interval and ${\bar A}$ its complement, are Gaussian. 
\label{e:d}
\end{enumerate}
Property \ref{e:b})  allows us to express the full density matrix in
the form (\emph{cf}. Eq.~\eqref{eq:RDMgen})
\bea
\label{eq:DM}
\rho(t)&=&\frac{1}{Z}\Big\{\sum_{\mathcal O_e}\Big[
{}_{\rm R}\langle\psi_t|\mathcal O_e|\psi_t\rangle_{\rm R}
+{}_{\rm NS}\langle\psi_t|\mathcal O_e|\psi_t\rangle_{\rm NS}\Big]
\mathcal O_e\nn
&&\quad +\sum_{\mathcal O_o}
2\mathrm{Re}\bigl[\!\!\!\!\!\!\!\!\!\!\!\!{\phantom{\braket{\mathcal
        O_o}}}_{\rm NS}\!\braket{\psi_t|\mathcal O_o|\psi_t}_{\rm
    R}\bigr]\mathcal O_o \Big\},
\eea
where $Z$ ensures that ${\rm Tr}[\rho(t)]=1$ and $\{\mathcal
O_e\}\cup\{\mathcal O_o\}$ is a complete set of Hermitian involutions
with the property  
\be\label{eq:algebra}
[e^{i\pi \mathcal N}, \mathcal O_e]=0,\qquad \{e^{i\pi \mathcal N}, \mathcal O_o\}=0\, .
\ee
We will refer to ${\cal O}_{e/o}$ as even and odd operators
respectively. The main difference between even and odd operators, is
that the latter are not local in terms of fermions: a Jordan-Wigner
string is attached to them. We are interested in the RDM of a block
$A$ of $\ell$ contiguous spins, which is obtained by tracing out the
degrees of freedom outside $A$ 
\be
\rho_{\ell}=\btr{\bar A}{\rho}\ .
\ee
A convenient representation for $\rho_\ell$ is obtained by restricting
the sums in Eq.~\eqref{eq:DM} to involutions that act as the identity
operator outside the interval $A$, \emph{i.e.} 
\be
\mathcal O\rightarrow\mathcal O^{(A)}\otimes \mathrm I^{(\bar A)}\, ,
\ee
where the superscript $(A)$ indicates that the operators act on the
Hilbert space over all sites in $A$. As a result of property
\ref{e:a}), fermion parity has a simple restriction onto the interval
$A$
\be
e^{i\pi \mathcal N_A}\equiv\prod_{l\in A}\sigma_l^z,
\ee
and can be used to subdivide operators ${\cal O}^{(A)}$ into even and
odd ones
\be\label{eq:algebraA}
[e^{i\pi \mathcal N_A}, \mathcal O_e^{(A)}]=0,\qquad \{e^{i\pi
\mathcal N_A}, \mathcal O_o^{(A)}\}=0\ .
\ee
This then implies that we can decompose the RDMs of \eqref{eq:DM} into
even and odd parts as well
\be
\rho_{\ell}=\rho_{\ell,e}+\rho_{\ell,o}\ .
\label{eq:bp}
\ee
In the thermodynamic limit we then may employ property \ref{e:c}) to
obtain the following expressions
\bea
\rho_{\ell,e}(t)&=&\frac{1}{2^\ell}\sum_{\mathcal O_e}
{}_{\rm R}\langle\psi_t|\mathcal O_e|\psi_t\rangle_{\rm R}\mathcal O_e\ ,\nn
\rho_{\ell,o}(t)&=&\frac{1}{2^\ell}\sum_{\mathcal O_o} 
\mathrm{Re}\bigl[{}_{\rm NS}\langle\psi_t|\mathcal
  O_o|\psi_t\rangle_{\rm R}\bigr]\mathcal O_o\, . 
\eea
Importantly, the even part $\rho_{\ell,e}(t)$ is
Gaussian~\eqref{eq:gaussian} by virtue of property \ref{e:d}), and has
the same structure as RDMs for quenches originating in the
paramagnetic phase. On the other hand, the odd part $\rho_{\ell,o}$
has its origin in the spontaneous breaking of the $\mathbb{Z}_2$
symmetry. The commutation relations \fr{eq:algebraA} imply that
$\btr{}{\rho_{\ell,o}(t)\rho^{\rm Ga}_\ell}=0$ for any
Gaussian density matrix $\rho^{\rm Ga}_\ell$, because the latter
is by construction even. As a result the odd part
$\rho_{\ell,o}$ of the RDM enters the distance from a Gaussian state
only through its norm
\be
\label{eq:Fdistance}
{\cal D}(\rho_\ell(t),\rho_\ell^{\rm Ga})=\sqrt{\frac{
\parallel\rho_{\ell,e}(t)-\rho_\ell^{\rm Ga}\parallel_F^2
+\parallel \rho_{\ell,o}(t)\parallel_F^2}
{\parallel\rho_{\ell,e}\parallel_F^2+\parallel\rho_{\ell,o}\parallel_F^2+
\parallel\rho_\ell^{\rm Ga}\parallel_F^2}
}.
\ee
We will be interested in the cases where $\rho^{\rm Ga}_\ell$
describe Gibbs or (truncated) generalized Gibbs ensembles. The Frobenius
norms $||\rho_{\ell,e}(t)-\rho_\ell^{\rm Ga}||_F$,
$||\rho_{\ell,e}||_F$ and $||\rho_\ell^{\rm
Ga}||_F$ can be efficiently evaluated by means of Eq.~\fr{eq:DFfree}. What remains in order to determine the distance
\fr{eq:Fdistance} is a method for calcuating the Frobenius norm
$\parallel\rho_{\ell,o}\parallel_F$. This is a somewhat involved
technical problem, which is addressed in Sec.~\ref{ss:cd} and Appendix~\ref{app1}.
The basic idea is to utilize a cluster decomposition theorem
at any finite time after the quench, see also
Ref.~[\onlinecite{EEF:2012}]. 

%%%%%%%%%%%%%%%%%%%%%%%%%%%%%%%%%%%%%%%%%%%%%%%%%%%%%%%%%%%%%%%%%%%%%%%%%%
\subsection{$\parallel\rho_{\ell,o}\parallel_F$ from cluster decomposition}\label{ss:cd}
\begin{figure}[t]
\includegraphics[width=0.47\textwidth]{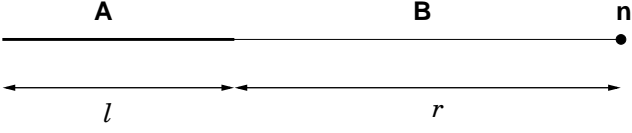}
\caption{Geometry of the composite system $A\cup n$ used in calculating
$\parallel\rho_{\ell,o}\parallel_F$, where $\rho_{\ell}$ is the RDM
of subsystem $A$. The single site at position $n$ is separated from
$A$ by a block $B$ of length $r$. }
\label{fig:SSB}
\end{figure}

%%%%%%%%%%%%%%%%%%%%%%%%%%%%%%%%%%%%%%%%%%%%%%%%%%%%%%%%%%%%%%%%%%%%%%%%%%
The main difficulty in calculating the Frobenius norm of
$\rho_{\ell,o}$ is that the latter is not Gaussian. The idea is
therefore to obtain $\rho_{\ell,o}$ as a reduction of a Gaussian
operator. To that end, we consider a composite system $C=A\cup n$
consisting of our subsystem $A$ and a single site at position $n$,
which is separated from $A$ by a block $B$ of length $r$, see
Fig.~\ref{fig:SSB}. 

The even part of the RDM $\rho_C(t)$ can be expanded in a complete
basis of Hermitian involutions ${\cal O}_{e/o}$ as
\bea
\label{eq:SSB0}
\rho_{C,e}(t)&=&
\frac{1}{2^{\ell+1}}\Bigl[\sum_{\mathcal O_e}\braket{\mathcal O_e\sigma^z_n}\mathcal
  O_e\sigma_n^z\nn
&&\quad+\sum_{\mathcal O_o}\sum_{\alpha=x,y}\braket{\mathcal
    O_o\sigma_n^{\alpha}}\mathcal O_o \sigma_n^\alpha\Bigr],
\eea
where $\braket{\cdots}=\braket{\Psi_t|\cdots|\Psi_t}\approx
{}_{\rm R}\langle\psi_t|\cdots|\psi_t\rangle_{\rm R}$, since both
${\cal O}_e\sigma^z_n$ and ${\cal O}_o\sigma^\alpha_n$ in
\eqref{eq:SSB0} are even operators with respect to fermion parity.
In the limit of large separation $r$, we may use the cluster
decomposition principle to simplify the expectation values
\bea
\braket{\mathcal O_e\sigma^z_n}&\xrightarrow{r\rightarrow\infty}&
\braket{\mathcal O_e}\braket{{\sigma^z_n}}\ ,\nn
\braket{\mathcal O_o\sigma^\alpha_n}&\xrightarrow{r\rightarrow\infty}&
\braket{\mathcal O_o}\braket{{\sigma^\alpha_n}}.
\eea
This then leads to the following relation between RDMs in the limit or
large separation 
\be
\lim_{r\to\infty}\rho_{C,e}(t)=\rho_{\ell,e}(t)\otimes\rho_{1,e}(t)
+\rho_{\ell,o}(t)\otimes\rho_{1,o}(t),
\label{decomp}
\ee
where $\rho_{1}$ is the RDM of site $n$. The piece of interest to us
is 
\be
\rho_{\ell,o}\otimes\rho_{1,o}=\lim_{r\to\infty}
\frac{1}{2^{\ell+1}}\sum_{\mathcal O_o}\sum_{\alpha=x,y}\braket{\mathcal
    O_o\sigma_n^{\alpha}}\mathcal O_o \sigma_n^\alpha.
\ee
In the next step we move from spins to Majorana fermions by means of
the Jordan-Wigner transformation \fr{eq:JW}
\be\label{eq:SSB1}
\rho_{\ell,o}\otimes\rho_{1,o}=\lim_{r\rightarrow\infty}\frac{1}{2^{\ell+1}}
\sum_{\ontop{\mathcal A_o}{\alpha=x,y}}\braket{a_n^\alpha \mathcal
  A_o^\dag e^{i\pi \mathcal N_{B}}}
\mathcal A_oe^{i\pi{\cal N}_B}a_n^\alpha,
\ee
where $\mathcal A_o$ are odd products of Majorana fermions acting on
sites within $A$. Importantly, the fermionic expression \fr{eq:SSB1} depends
on the configuration of Majoranas in subsystem $B$ through the
Jordan-Wigner string operator. The right hand side of \fr{eq:SSB1} can
be cast in the form
\be
\label{eq:SSB2}
\rho_{\ell,o}\otimes \rho_{1,o}=\lim_{r\rightarrow\infty}
\braket{e^{i\pi \mathcal N_{B}}}e^{i\pi \mathcal
  N_{B}}\frac{\text{\textgoth p}-\sigma_n^z\text{\textgoth
p}\sigma_n^z}{2}\ ,
\ee
where $\textgoth{p}$ is a normalized, Gaussian operator
\fr{eq:gaussian} acting on the Hilbert space over sites $A\cup n$ 
\be
\text{\textgoth p}\equiv\frac
{\tr{\overline{A\cup n}}{e^{i\pi \mathcal N_{B}} |\psi_t\rangle_{\rm
      R}{}_{\rm R}\langle\psi_t|}}{\braket{e^{i\pi \mathcal N_{B}}}}\, .
\ee
In writing \eqref{eq:SSB1} we are assuming $\braket{e^{i\pi \mathcal
    N_{B}}}\neq 0$. The fact that $\textgoth{p}$ is Gaussian is a
consequence of the particular form of $\ket{\psi_t}_{\rm R}$ (which is
the analog of \fr{Psit_para} in the R sector) and $\mathcal N_B$ being
quadratic in fermions. The odd part of the single-site RDM is of the form
\be
\rho_{1,o}(t)=m^x(t)\sigma^x_n+m^y(t)\sigma^y_n\ ,
\ee
and hence 
\be
[\rho_{1,o}(t)]^2=\big([m^x(t)]^2+[m_y(t)]^2\big) I_2\equiv m_\perp^2(t)I_2.
\label{r1o}
\ee
Here the late-time behaviour of $m^{x,y}(t)$ are given by \fr{eq:mx}
and \fr{eq:my} respectively, and following Ref.~[\onlinecite{CEF1:2012}] they can
be easily calculated numerically for all times.
Combining \fr{r1o} and \fr{eq:SSB2} we obtain
\bea
\label{eq:F0}
\parallel\rho_{\ell,o}\parallel_F&=&
\lim_{r\rightarrow\infty}\frac{|\braket{e^{i\pi \mathcal
      N_{B}}}|}{\sqrt{2} |m_\perp(t)|}\Big|\hskip-0.8pt\Big|\frac{{\textgoth
  p}-\sigma_n^z{\textgoth p}\sigma_n^z}{2}\Big|\hskip-0.8pt\Big|_F\nn
&=&\lim_{r\rightarrow\infty}
\frac{|\braket{e^{i\pi \mathcal N_{B}}}|}{2 |m_\perp(t)|}
\sqrt{\btr{}{\text{\textgoth p}^2-(\sigma_n^z\text{\textgoth p})^2}}.\nn
\eea
Since both $\text{\textgoth p}$ and $\sigma_n^z\text{\textgoth
  p}\sigma_n^z$ are Gaussian, their moments can be written in terms of
their respective correlation matrices 
\bea\label{eq:GG}
\mathcal G_{i j}&\equiv& \tr{}{\text{\textgoth p}a_j a_i}-\delta_{i
  j}\ ,\nn
\bar{\mathcal G}_{i j}&\equiv& \tr{}{\sigma_n^z\text{\textgoth
    p}\sigma_n^z a_j a_i}-\delta_{i j}\, . 
\eea
We note that the correlation matrices are related by $\bar{\mathcal
  G}=P_n \mathcal G P_n$, with $P_n$ the diagonal matrix that changes
the sign of the last 2-by-2 block ($\mathrm I_d$ is the $d\times d $
identity) \be\label{eq:Pn}
P_n=\mathrm I_{2\ell} \oplus (-\mathrm I_{2})\, .
\ee
Using \eqref{eq:trbasic} we have
\be
\btr{}{\text{\textgoth p}^2}=\{\mathcal G,\mathcal
G\}\, ,\quad \btr{}{(\sigma_n^z \text{\textgoth
    p})^2}=\{\mathcal G,\bar{\mathcal G}\}\, . 
\ee 
A slight complication arises because \textgoth{p} is not positive
semidefinite. To account for this we must use the more general
definition of $\{\Gamma,\Gamma^\prime\}$ as the product of the
eigenvalues of $(1+\Gamma\Gamma^\prime)/2$ with halved
degeneracy~\cite{FC:2010}.  We may then recast \eqref{eq:F0} in the form
\be
\label{eq:SSB4.5}
\parallel
\rho_{\ell,o}\parallel_F=\lim_{r\rightarrow\infty}\frac{|\braket{e^{i\pi
      \mathcal  N_{B}}}|}{2|m_\perp(t)|}\sqrt{\{\mathcal G,\mathcal
  G\}-\{\mathcal G,\bar{\mathcal G}\}}\, . 
\ee
While formally correct, \fr{eq:SSB4.5} is not suitable for numerical
computations, because at large distances $\braket{e^{i\pi \mathcal
    N_{B}}}$ becomes very close to zero. A more convenient expression
derived in Appendix \ref{app1} is
\be
\label{eq:SSB5}
\parallel \rho_{\ell,o}\parallel_F=\lim_{r\rightarrow\infty}
\frac{\sqrt{\det\bigl(\mathrm I_{2\ell}\oplus 0_{2r}\oplus\mathrm
    I_2+i\Gamma_{A\cup B\cup n}\bigr)}}{2^{1+\ell/2}|m_\perp(t)|}\,
. 
\ee
Here $\Gamma_{A\cup B\cup n}$ is the correlation matrix of the
interval $A\cup B\cup n$ and is given by \fr{eq:Gamma},
\fr{eq:Gamma_l}, \fr{eq:fg}. In order to utilize \fr{eq:SSB5} we in
principle have to consider infinite separations $r$ and hence
infinitely large matrices. 

\begin{figure}[t]
\includegraphics[width=0.47\textwidth]{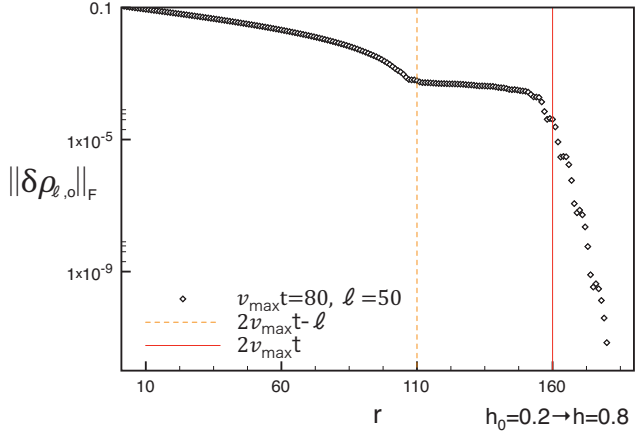}
\caption{Difference  $\displaystyle
\parallel\delta\rho_{\ell,o}\parallel_F=\parallel\rho_{\ell,o}\parallel_F
-\lim_{r\to\infty}\parallel\rho_{\ell,o}\parallel_F$ as a function of
the separation $r$ for a quench from $h_0=0.2$ to $h=0.8$. We see that
for $r>2v_{\rm max}t$ the difference becomes exponentially small in
$r/\xi$, where $\xi$ is the correlation length in the initial state.
  } 
\label{fig:SSB_norm}
\end{figure}

Crucially, in practice a \emph{finite}
separation $r>2v_{\rm max}t$, where $v_{\rm max}={\rm
  max}_k\epsilon_h(k)$ is the maximal propagation velocity, is
sufficient to recover the $r\to\infty$ limit up to corrections that
are exponentially small in $r/\xi$. Here $\xi$ is the correlation
length in the initial state. A representative example is shown in
Fig.~\ref{fig:SSB_norm}. 
In practice, using a finite $r>2v_{\rm max}t+\xi\delta$ with
$\delta\approx 20$ provides an efficient way for calculating
$||\rho_{\ell,o}(t)||_F$ and then by means of \fr{eq:Fdistance}
distances ${\cal D}(\rho_\ell(t),\rho_\ell^{\rm Ga})$ for quenches
originating in the ferromagnetic phase.
%%%%%%%%%%%%%%%%%%%%%%%%%%%%%%%%%%%%%%%%%%%%%%%%%%%%%%%%%%%%%%%%%%%%%%%%%%
\subsection{Results for quenches from the ferromagnetic phase}
%%%%%%%%%%%%%%%%%%%%%%%%%%%%%%%%%%%%%%%%%%%%%%%%%%%%%%%%%%%%%%%%%%%%%%%%%%

For quenches with $h_0<1$ and in the thermodynamic limit, we determine
the distance between the quench RDM and that of an appropriate thermal
or generalized Gibbs ensemble by means of relations \fr{eq:Fdistance}
and \fr{eq:SSB5}. The correlation matrices for all cases are of the
form \fr{eq:Gamma}, \fr{eq:Gamma_l} with elements given in
\fr{eq:thermal}, \fr{eq:gGGE} and \fr{eq:fg} respectively. For a
subsystem of size $\ell$ most terms require the calculation of
determinants of $2\ell\times 2\ell$ matrices, which is easily done
numerically. The evaluation of $\parallel\rho_{\ell,0}\parallel_F$ is
significantly more costly, and in practice involves
determinants of at most $2(\ell+2v_{\rm \max}t+\xi\delta)\times
2(\ell+2v_{\rm max}t+\xi\delta)$ matrices, as discussed above.

\begin{figure}[ht]
\begin{center}
\includegraphics[width=0.45\textwidth]{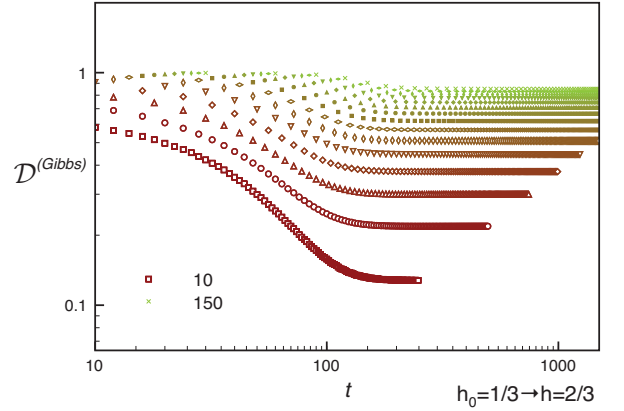}
\caption{
Distance 
${\mathcal D}^{\rm Gibbs}=
\mathcal D(\rho_\ell(t),\rho^{\beta}_\ell)$ 
after a quench within the ferromagnetic phase for subsystem sizes
$\ell=10,20,\dots,150$. As $\ell$ increases, the color fades from
brown to green, the symbols become smaller and the curves narrower. 
At late times the distances tend to constants depending on subsystem
size. 
}\label{fig:ferro_gibbs}
\end{center}
\end{figure}

\begin{figure}[ht]
\begin{center}
\includegraphics[width=0.45\textwidth]{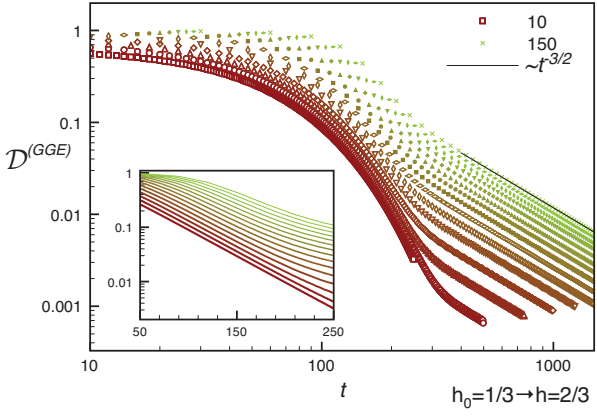}
\caption{Distance 
${\mathcal D}^{\rm GGE}=\mathcal D(\rho_\ell(t),\rho^{GGE}_\ell)$, after a
quench within the ferromagnetic phase for the subsystem lengths
$\ell=10,20,\dots,150$. We used the same notations of
Fig.~\ref{fig:para_gge}. The behavior is almost the same as that shown
in Figs~\ref{fig:para_gge} and \ref{fig:chipara}, but the effect of
the spontaneous magnetization is visible at intermediate times, when
the distance decays exponentially (inset). 
}\label{fig:chiferro}
\end{center}
\end{figure}

\begin{figure}[ht]
\begin{center}
\includegraphics[width=0.45\textwidth]{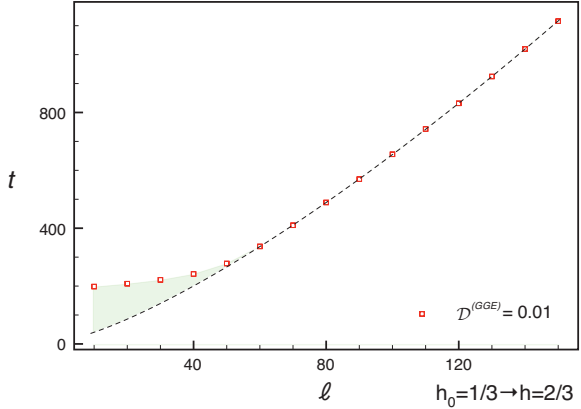}
\caption{The time \emph{vs.} the subsystem length at fixed distance
$\mathcal D(\rho_\ell(t),\rho^{GGE}_\ell)=0.01$ (black solid line of
the left plot). The dashed curve is $t=a+b\ell^{4/3}$, with $a$ and
$b$ obtained by fitting the numerical data. The filled region shows
the effect of the spontaneous magnetization.  
}\label{fig:chiferro_2}
\end{center}
\end{figure}

Results for a quench from $h_0=1/3$ to $h=2/3$ and subsystem sizes
$\ell=10,20,30\ldots,150$ are  
shown in Figs~\ref{fig:ferro_gibbs} and \ref{fig:chiferro}.
We see that the distance between quench and Gibbs RDMs tends to a
$\ell$-dependent constant at late times. On the other hand, as shown in
Fig.~\ref{fig:chiferro}, at sufficiently late times the distance
between $\rho_\ell(t)$ and $\rho_{\rm GGE,\ell}$ decays to zero in a 
universal power-law fashion
\be\label{eq:Dnum_ferro}
\mathcal D(\rho_\ell(t),\rho_{\rm GGE,\ell})\xrightarrow{J t\gg
  1}k(\ell)(J t)^{-3/2}+\dots\, .
\ee
The large-$\ell$ asymptotics of the function $k(\ell)$ can be inferred
in the same way as for quenches within the paramagnetic phase.
On surfaces with constant, small $\mathcal D$, time scales
as $t\sim \ell^{4/3}$ as is shown in Fig.~\ref{fig:chiferro_2}, which
implies that
\be
k(\ell)\sim \ell^{2}\, .
\ee
We conclude that the late time behaviour of the distance between 
quench and generalized Gibbs RDMs is the same as for quenches
within the paramagnetic phase. The mean relaxation time $t^*_{\rm
  rms}$ is therefore again given by \eqref{eq:relaxrms}. Interestingly
this coincides with the result obtained in Ref.~[\onlinecite{CEF2:2012}]
for the relaxation of the order parameter two-point function after
quenches within the ferromagnetic phase. The effects of the
spontaneous symmetry breaking are important only at short and
intermediate times. It is shown in the inset of
Fig.~\ref{fig:chiferro} that there is a time window, in which the odd
part of the RDM gives the dominant contribution to the distance, which
decays exponentially. 

%%%%%%%%%%%%%%%%%%%%%%%%%%%%%%%%%%%%%%%%%%%%%%%%%%%%%%%%%%%%%%%%
\subsection{Magnitude of the contribution due to $\rho_{\ell,o}$}
%%%%%%%%%%%%%%%%%%%%%%%%%%%%%%%%%%%%%%%%%%%%%%%%%%%%%%%%%%%%%%%%
The effects of the spontaneously broken $\mathbb{Z}_2$ symmetry in the
initial state make themselves felt through the $\mathbb{Z}_2$-odd part
$\rho_{\ell,o}$ of the density matrix. The relative importance of
$\rho_{\ell,o}$ for large $\ell$ can be estimated by considering the
von Neumann entropy of subsystem $A$
\be
S_{\rm vN}[\rho_{\ell}]={\rm
  Tr}\left[\rho_{\ell}\ln\left(\rho_{\ell}\right)\right]. 
\ee
We recall that the von Neumann entropy after a global quench grows
linearly in time until the Fermi time $t_F=\ell/(2 v_{\rm max})$, and
then saturates to a value proportional to the subsystem size $\ell$
\cite{CC:2005,FC:2008}. 
Using the commutation relations \eqref{eq:bp} we see that
the even part $\rho_{\ell,e}$ can be expressed in terms of the full
RDM $\rho_\ell$ as follows 
\be\label{eq:free}
\rho_{\ell,e}=\frac{\rho_{\ell}+e^{i\pi \mathcal
    N_A}\rho_{\ell}e^{i\pi \mathcal N_A}}{2}\, . 
\ee
Since for any set of density matrices $\rho_i$ the von Neumann entropy
satisfies\cite{W:1978} ($\lambda_i>0$, $\sum_i\lambda_i=1$) 
\be
\sum_i \lambda_i\log\lambda_i\leq S_{\rm vN}\bigl[\sum_i\lambda_i
  \rho_i\bigr]-\sum_i\lambda_iS_{\rm vN}\bigl[\rho_i\bigr]\leq 0\, , 
\ee
the following bounds on the von Neumann entropy of subsystem $A$ hold
\be\label{eq:SvNbounds}
S_{\rm vN}[\rho_{\ell,e}]-\log 2\leq S_{\rm vN}[\rho_{\ell}]\leq
S_{\rm vN}[\rho_{\ell,e}]. 
\ee 
This demonstrates that at any time after the quench the symmetry
breaking contribution to the von Neumann entropy will be at most
$\log 2$. Given that for large subsystems the von Neumann entropy at
late times is proportional to $\ell$, we conclude that the relative
contribution of the odd part of the RDM will be important only for
small subsystem sizes.

%%%%%%%%%%%%%%%%%%%%%%%%%%%%%%%%%%%%%%%%%%%
\subsubsection{A conjecture for $\parallel\rho_{\ell,o}\parallel_F$
in the limit of large $\ell$ and $Jt$
} 
\begin{figure}[ht]
\begin{center}
\includegraphics[width=0.45\textwidth]{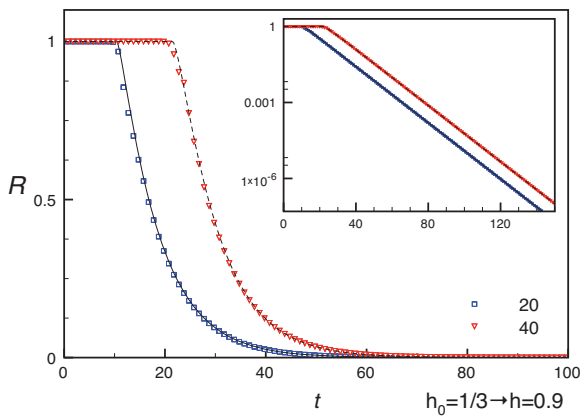}
\caption{The ratio
  $R=\frac{\parallel\rho_{\ell,o}(t)\parallel_F}
{\parallel\rho_{\ell,e}(t)\parallel_F}$ after a quench within the
ferromagnetic phase for subsystem lengths $\ell=10,20$. The lines
correspond to the analytic expression~\eqref{eq:CONJ}, where we have
included a correction ${\cal O}(\ell^0)$ by shifting
$\ell\rightarrow\ell-1.2$. The inset presents the same data on a 
logarithmic scale.
}\label{fig:odd}
\end{center}
\end{figure}

%%%%%%%%%%%%%%%%%%%%%%%%%%%%%%%%%%%%%%%%%%%
We now consider the \emph{space-time scaling limit}\cite{CEF1:2012}
\be
\ell,Jt\to\infty\ ,\quad \frac{\ell}{Jt}\ {\rm fixed}.
\ee
We observe that in this limit our numerical results for quenches
within the ferromagnetic phase are in excellent agreement with the
following relation
\begin{widetext}
\be
\label{eq:CONJ}
\log \parallel\rho_{\ell,o}(t)\parallel_F\approx \log
\parallel\rho_{\ell,e}(t)\parallel_F+\int_0^\pi\frac{\mathrm d
  k}{2\pi}\log\big(\cos\Delta_k\big) 
\max_k\big\{0,2\varepsilon^\prime_k   t-  \ell+\mathcal O(\ell^0,t^0)\big\}\, .
\ee
\end{widetext}
Here we have highlighted the asymptotic nature of the relation and
indicated by $\mathcal O(\ell^0,t^0)$, where the most important
corrections will arise. Since $\log ||\rho_{\ell,e}(t)||_F$ is
proportional to the R\'enyi entropy $S_2$ (\emph{cf.}
Eq.~\eqref{eq:S2blah}), we may use the known results \cite{FC:2008} on
the asymptotics of the latter
\begin{multline}
\log \parallel\rho_{\ell,e}(t)\parallel_F= - S_2/2 \approx\\
\int_0^\pi\frac{\mathrm d k}{2\pi}\log\frac{1+\cos^2\Delta_k}{2}\min(2\varepsilon^\prime_k t,\ell)+\mathcal O(\ell^0,t^0)\, .
\label{normrhoe}
\end{multline}
Combining \fr{normrhoe} and \fr{eq:CONJ} provides a conjecture for the
asymptotic behaviour of $\parallel\rho_{\ell,o}\parallel_F$. This
conjecture is compared to numerical results in Fig.~\ref{fig:odd}.
The agreement is clearly quite good.

%%%%%%%%%%%%%%%%%%%%%%%%%%%%%%%%%%%%%%%%%%%%
\section{Quenches across the critical point}%
\begin{figure}[ht]
\begin{center}
\includegraphics[width=0.45\textwidth]{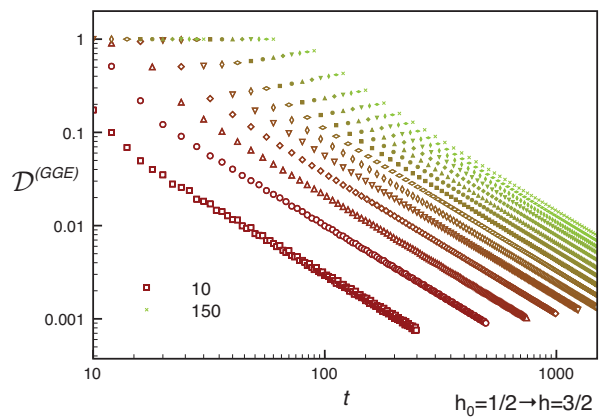}
\caption{Distance 
${\mathcal D}^{\rm GGE}=
\mathcal D(\rho_\ell(t),\rho^{GGE}_\ell)$, after a 
quench from ferromagnetic phase to the paramagnetic phase for the
subsystem lengths $\ell=10,20,\dots,150$. The conventions are the same
as in Fig.~\ref{fig:para_gge}.  
}\label{fig:chiferropara}
\end{center}
\end{figure}

\label{sec:across}
%%%%%%%%%%%%%%%%%%%%%%%%%%%%%%%%%%%%%%%%%%%%
We now turn to quenches across the critical point. These are of
particular interest \cite{rsms-08,CEF1:2012,heyl}. In
Fig.~\ref{fig:chiferropara} we plot the distance between quench and
GGE reduced density matrices for a quench from the ferromagnetic phase
($h_0=1/2$) to the paramagnetic phase ($h=3/2$). The 15 data sets
displayed correspond to subsystem sizes between $\ell=10$ and
$\ell=150$. We find that the distance
${\mathcal D}^{\rm GGE}=\mathcal D(\rho_\ell(t),\rho^{GGE}_\ell)$
again decays in a universal $t^{-3/2}$ power law.
In Fig.~\ref{fig:chiferropara} we consider the same quench, but focus
on very small subsystem sizes $\ell=1,2,3,4$. We observe that the
distance displays an oscillatory behaviour on top of a power-law decay
in time. This is in agreement with the analytic results discussed in
section \ref{s:ferro1} for the $\ell=1$ case. Increasing the subsystem
size leads to a rapid suppression of the amplitude of the oscillations.

\begin{figure}[ht]
\begin{center}
\includegraphics[width=0.45\textwidth]{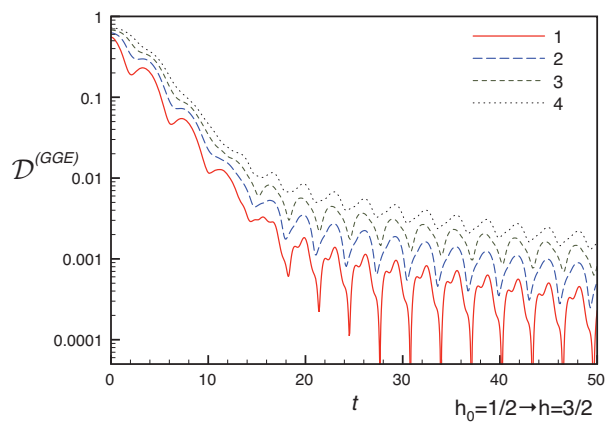}
\caption{Distance $
{\mathcal D}^{\rm GGE}=\mathcal D(\rho_\ell(t),\rho^{GGE}_\ell)$, after a
quench from ferromagnetic phase to the paramagnetic phase for the small subsystems 
$\ell=1,2,3,4$.
}\label{fig:chiferroparasmall}
\end{center}
\end{figure}

\begin{figure}[ht]
\begin{center}
\includegraphics[width=0.45\textwidth]{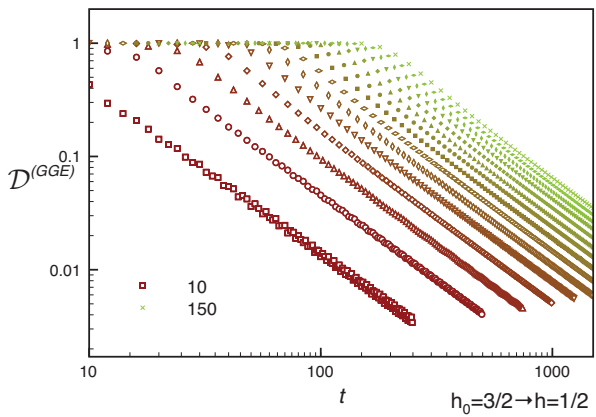}
\caption{Distance $\mathcal D(\rho_\ell(t),\rho^{GGE}_\ell)$, after a
quench from paramagnetic phase to the ferromagnetic phase for the subsystem lengths
$\ell=10,20,\dots,150$. The conventions are the same as in
Fig.~\ref{fig:para_gge}. 
}\label{fig:chiparaferro}
\end{center}
\end{figure}

\begin{figure}[ht]
\begin{center}
\includegraphics[width=0.45\textwidth]{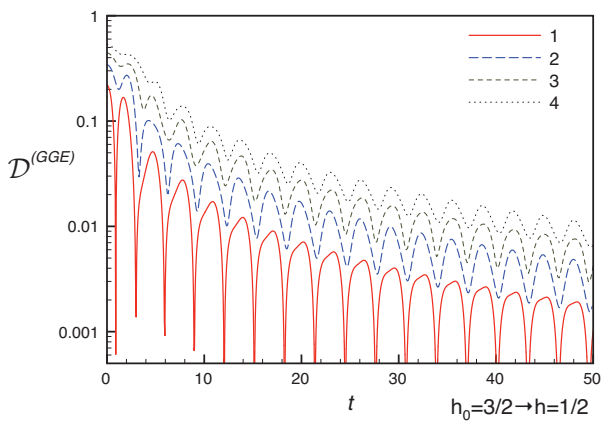}
\caption{Distance $\mathcal D(\rho_\ell(t),\rho^{GGE}_\ell)$, after a
quench from paramagnetic phase to the ferromagnetic phase for the small subsystems 
$\ell=1,2,3,4$.
}\label{fig:chiparaferrosmall}
\end{center}
\end{figure}

In Figs~\ref{fig:chiparaferro} and \ref{fig:chiparaferrosmall} we
consider the reverse quenches, i.e. starting at $h_0=3/2$ in the
paramagnetic phase, and quenching to $h=1/2$ in the ferromagnetic
phase. The behaviour of the distances is very similar to what we found
for the quench from $h_0=1/2$ to $h=3/2$: at late times the distance
decays as a $t^{-3/2}$ power law, and for small subsystem sizes we
observe oscillatory behaviour on top of this decay.

%%%%%%%%%%%%%%%%%%%%%%%%%%%%%%%%%%%%%%%%%%
\section{Summary and Conclusions}
\label{sec:summary}
%%%%%%%%%%%%%%%%%%%%%%%%%%%%%%%%%%%%%%%%%%
In this work we have considered the evolution of reduced density
matrices after a quantum quench in the transverse field Ising chain.
The main result of our work is to demonstrate that 
\be
\lim_{t\to\infty}\rho_\ell(t)=\rho_{\rm GGE,\ell},
\label{summ1}
\ee
where $\rho_\ell(t)$ is the reduced density matrix of a subsystem
consisting of $\ell$ adjacent spins after a quench of the transverse
field, and $\rho_{\rm GGE,\ell}$ is the reduced density matrix of an
appropriately defined generalized Gibbs ensemble. The derivation of
\fr{summ1} is based on defining an appropriate distance ${\cal
  D}(\rho,\rho')$ on the space of reduced density matrices, and then
establishing that the distance between quench and GGE reduced density
matrices approaches zero at late times. For our particular choice of
distance we found that at late times this distance approaches zero as
a universal power law in time
\be
{\cal D}(\rho_\ell(t),\rho_{\rm GGE,\ell})\sim t^{-3/2}\ .
\ee

We have presented a
detailed construction of $\rho_{\rm GGE,\ell}$ in terms of the
\emph{local} (in space) integrals of motion $I^\pm_n$ of the
TFIC. The densities of these conservation laws involve only spins on
$n+2$ consecutive sites. We
proved that these local conservation laws are related in a
\emph{linear} fashion to the occupation numbers of the Bogoliubov
fermions that diagonalize the Hamiltonian of the TFIC. This linear
relation establishes the equivalence of our construction of the GGE to
the one frequenctly used in the literature, which is based on mode
occupation numbers.

We then have addressed the question, which of the conservation laws
are most important for obtaining an accurate description of the
stationary limit $\lim_{t\to\infty}\rho_{\ell}(t)$ of the quench RDM.
To that end we introduced (defective) truncated generalized Gibbs
ensembes, which are missing some of the local conservation laws. We
found that the more local the conservation laws (i.e. the fewer
consecutive spins their densities involve), the more important they
are for describing the stationary state of a given subsystem. Loosely
speaking we observed that in order to obtain a good description of the
stationary state RDM of a subsystem of size $\ell$, we need to retain
all local conservation laws, whose densities involve at most $\approx
\ell+n_0$ neighbouring spins, where $n_0$ is a constant depending on
$h_0$ and $h$. Leaving out ``highly local'' conservation laws
generally provides a very poor description of the stationary state.
To the best of our knowledge this is the first such demonstration of a
connection between locality of conservation laws and their importance
in the GGE context.

Our work raises a number of issues. First and foremost is the 
dependence of the results obtained on the precise definition of the
distance on the space of reduced density matrices. We have argued,
that the ``best'' distance is the one based on the trace norm, because
it provides the most direct and precise information on the time
evolution of local observables. Unfortunately this distance is much
harder to handle analytically. It would however be very interesting to
implement it in purely numerical studies using iTEBD or related
algorithms. 

\acknowledgments
We thank P. Calabrese and J. Eisert for helpful discussions.
This work was supported by the EPSRC under grants EP/I032487/1 and
EP/J014885/1 and the National Science Foundation under grant NSF
PHY11-25915 (FHLE). FHLE is grateful to the KITP in Santa Barbara for
hospitality. 

\appendix

%%%%%%%%%%%%%%%%%%%%%%%%%%%%%%%%%%%%%%%%%%%%%%%%%%%%%%%%
\section{Inequalities involving the Frobenius norm of RDMs for spin-1/2 Quantum Spin Chains}\label{app:frob}%
%%%%%%%%%%%%%%%%%%%%%%%%%%%%%%%%%%%%%%%%%%%%%%%%%%%%%%%%
In this appendix we provide lower and upper bounds for the Frobenius
norm of the difference of two reduced density matrices
$\parallel\rho-\rho'\parallel_F$ in a translationally invariant
system. An upper bound is obtained as follows
\bea
\parallel\rho-\rho'\parallel_F^2&=&{\rm
  Tr}\left[\rho^2+\rho'^2-2\rho\rho'\right]\nn
&=&\parallel\rho\parallel_F^2+\parallel\rho'\parallel_F^2
-2{\rm  Tr}\left(\rho\rho'\right)\nn
&\leq&\parallel\rho\parallel_F^2+\parallel\rho'\parallel_F^2.
\label{eq:upper_bound}
\eea
Here we have used that both $\rho$ and $\rho'$ are positive semidefinite
and hence
\bea
{\rm
  Tr}\left(\rho\rho'\right)&=&\sum\lambda_j\rho'_{jj}\geq\lambda_{\rm
  min}\sum_j\rho'_{jj}\nn
&=&
\lambda_{\rm  min}{\rm Tr}\rho'=\lambda_{\rm min}\geq 0,
\eea
where $0\leq \lambda_{\rm min}\leq\lambda_j$ are the eigenvalues of $\rho$.
To derive a lower bound we start by expressing the RDM of a block
of  $\ell$ spins in a spin-$\frac{1}{2}$ chain in the form
\be
\rho_\ell=\frac{1}{2^\ell}\sum_{\{\alpha_j\}}\tr{}{\rho\,
  \sigma_{1}^{\alpha_1}\cdots\sigma_{\ell}^{\alpha_\ell}}\sigma_{1}^{\alpha_1}\cdots\sigma_{\ell}^{\alpha_\ell}\,
, 
\ee
where $\alpha_i=0,x,y,z$
with $\sigma^{0}\equiv \mathrm I$, and $\rho$ is the density matrix of
the full system;  $\rho_\ell$ is only function of the length because of translational invariance.
By singling out the term with $\alpha_\ell=0$, we can
express this in the form 
\be\label{eq:rhoell}
\rho_\ell=\frac{\rho_{\ell-1}\otimes \mathrm I}{2}+\sum_{\alpha_\ell=1}^3\delta\rho_{\ell-1}^{\alpha_\ell}
\sigma_{\ell}^{\alpha_\ell}\ ,
\ee
where $\rho_{\ell-1}$ is the RDM of the block consisting of sites
$1,\dots, \ell-1$.
We also write the RDM of the $\ell^{th}$ spin
\be
\rho_1=\frac{\mathrm I}{2}+\sum_{\alpha_\ell=1}^3\tr{}{\delta\rho_{\ell-1}^{\alpha_\ell}}
\sigma_{\ell}^{\alpha_\ell}
\ee
and observe that
\be\label{eq:aux1a}
\parallel\rho_1-\rho_1^\prime\parallel_F^2=2\sum_{\alpha_\ell=1}^3
\left(\tr{}{\Omega^{\alpha_\ell}_{\ell-1}}\right)^2\, . 
\ee
Here we have defined $\Omega^{\alpha_\ell}_{\ell-1}=\delta\rho_{\ell-1}^{\alpha_\ell}
-\delta\rho'^{\alpha_\ell}_{\ell-1}$.
Using \eqref{eq:rhoell} we have
\bea
\label{eq:A5}
\parallel\rho_\ell-\rho'_\ell\parallel_F^2&=&
\frac{\parallel\rho_{\ell-1}-\rho'_{\ell-1}\parallel_F^2}{2}
+2\sum_{\alpha_\ell=1}^3\parallel\Omega^{\alpha_\ell}_{\ell-1}\parallel_F^2\nn
&\geq&\frac{\parallel\rho_{\ell-1}-\rho'_{\ell-1}\parallel_F^2}{2}
+\sum_{\alpha_\ell=1}^3
\frac{\left(\tr{}{\Omega^{\alpha_\ell}_{\ell-1}}\right)^2}{2^{\ell-2}}\nn
&=&\frac{\parallel\rho_{\ell-1}-\rho'_{\ell-1}\parallel_F^2}{2}+
\frac{\parallel\rho_{1}-\rho'_{1}\parallel_F^2}{2^{\ell-1}}\, ,
\eea
where we have used that for $N\times N$ matrices $M$ we have $N{\rm
  Tr} M^2\geq ({\rm Tr} M)^2$ in the second step, and \fr{eq:aux1a} in
the last. Iterating Eq.~\eqref{eq:A5} $\ell-1$ times we obtain
\be
\parallel\rho_\ell-\rho'_\ell\parallel_F^2\geq
2^{1-\ell}\ell\parallel\rho_1-\rho'_1\parallel_F^2.
\ee
This implies that for sufficiently large subsystem size $\ell$, the
distance $\parallel\rho_\ell-\rho'_\ell\parallel_F$ will generally be
larger than $2^{1-\ell/2}$.

%%%%%%%%%%%%%%%%%%%%
\section{Derivation of Eq.  \fr{eq:SSB5}}
\label{app1} %
%%%%%%%%%%%%%%%%%%%%
Our starting point is Eq.~\eqref{eq:SSB4.5}, i.e.
\be
\label{eq:SSB4.5app}
\parallel
\rho_{\ell,o}\parallel_F=\lim_{r\rightarrow\infty}\frac{|\braket{e^{i\pi
      \mathcal  N_{B}}}|}{2|m_\perp(t)|}\sqrt{\{\mathcal G,\mathcal
  G\}-\{\mathcal G,\bar{\mathcal G}\}}\, . 
\ee
Our task is to evaluate
\be\label{eq:tosimp}
\braket{e^{i \pi\mathcal N_B}}^2\{\mathcal G,\mathcal G\}\quad \text{and}\quad \braket{e^{i \pi\mathcal N_B}}^2\{\mathcal G,\overline{\mathcal G}\}\, ,
\ee
where $\overline{\mathcal G}=P_n\mathcal G P_n$ and $P_n$ is the diagonal
involution defined in \eqref{eq:Pn}. 
We recall that
$\{Q,Q_1\}$ denotes the product of the eigenvalues of
$(\mathrm I+ Q Q_1)/2$ with halved degeneracy (the eigenvalues of 
$QQ_1$ are always double degenerate~\cite{IF:2009} for
antisymmetric matrices $Q$ and $Q_1$). 
The correlation matrix $\mathcal G$ defined in Eq.~\eqref{eq:GG} turns out to be~\cite{FC:2010} the Schur complement of the block matrix $\Gamma_{B}$ of the matrix $\Gamma_{A\cup B\cup n}$, \emph{i.e.}
\be\label{eq:Gcal}
\mathcal G=\Gamma_{A\cup n}-\Gamma_{A\cup n,\, B}\frac{1}{\Gamma_{B}}\Gamma_{B,\, A\cup n}\, .
\ee
Here $\Gamma_{R_1,\, R_2}$ denotes the matrix, whose rows and columns
are associated with spatial regions $R_1$ and $R_2$ respectively,
\emph{e.g.} 
\be
[\Gamma_{A\cup n,\, B}]_{i j}=\begin{cases}
\Gamma_{i,j+2\ell}&0\leq i\leq 2\ell\\
\Gamma_{i+2r,j+2\ell}&i>2\ell\, .
\end{cases}
\ee
Here $\ell$ is the size of subsystem $A$, while $r$ is the distance
between $A$ and site $n$, see Fig.~\ref{fig:SSB}.

The first step is to find determinant representations for $\{Q,Q\}$
and $\{Q,PQP\}$, where $P$ is a generic symmetric involution ($P^2=\mathrm
I$ and $P^t=P$). 

We first consider $\{Q,Q\}$. Since $Q$ is antisymmetric, its eigenvalues
come in pairs $\pm q$
\be
0=\det|Q-q\mathrm I|=\det|Q^t-q\mathrm I|=\det|-Q-q\mathrm I|\, .
\ee
Both eigenvalues $\pm q$ give rise to the same eigenvalue $1+q^2$ of
$\mathrm I+Q^2$, and hence
\be
\{Q,Q\}=\prod_{q>0}\frac{1+q^2}{2}\, .
\ee
Here the product is over all positive eigenvalues of $Q$.
Using that
\be
\det|\mathrm I+i Q|=\prod_{q_>0}(1+iq)\prod_{q_>0}(1-iq)=\prod_{q>0}(1+q^2)\, ,
\ee
it follows that
\be\label{eq:aux1}
\{Q,Q\}=\det|(\mathrm I+i Q)/\sqrt{2}|\, .
\ee
Next we consider $\{Q,PQP\}$. The matrix
\be
P^{\frac{1}{2}}\equiv \frac{e^{i\pi/4}\rm I+e^{-i\pi/4}P}{\sqrt{2}}
\ee
satisfies $(P^{\frac{1}{2}})^2=P$ and
$(P^{\frac{1}{2}})^t=P^{\frac{1}{2}}$. 
Since we have
\be
\mathrm I+QPQP=(P^{\frac{1}{2}})^{-1}(\mathrm I+(P^{\frac{1}{2}}QP^{\frac{1}{2}})^2)P^{\frac{1}{2}}\, ,
\ee
the eigenvalues of $\mathrm I+QPQP$ and $\mathrm
I+(P^{\frac{1}{2}}QP^{\frac{1}{2}})^2$ coincide. Therefore
\be\label{eq:aux20}
\{Q,PQP\}=\{P^{\frac{1}{2}}QP^{\frac{1}{2}},P^{\frac{1}{2}}QP^{\frac{1}{2}}\}=\det\Bigl|\frac{\mathrm I+i P^{\frac{1}{2}}QP^{\frac{1}{2}}}{\sqrt{2}}\Bigr|\, ,
\ee
where in the last step we used Eq.~\eqref{eq:aux1}. Since
$(P^{\frac{1}{2}})^2= P$ and $P^2=\mathrm I$, \fr{eq:aux20} can be
rewritten in the form
\be\label{eq:aux2}
\{Q,PQP\}=\det|P|\det|(P+iQ)/\sqrt{2}|\, .
\ee
Using \eqref{eq:aux1} and \eqref{eq:aux2}, we can reexpress the
quantities in \eqref{eq:tosimp} as follows:  
\be\label{eq:OK0}
\ba
\braket{e^{i \pi\mathcal N_B}}^2\{\mathcal G,\mathcal G\}&=\det|i\Gamma_B| \det|(\mathrm I+i \mathcal G)/\sqrt{2}|\\
\braket{e^{i \pi\mathcal N_B}}^2\{\mathcal G,\overline{\mathcal G}\}&=\det|i\Gamma_B| \det|(P_n+i \mathcal G)/\sqrt{2}|\, .
\ea
\ee
Here we have use that the expectation value of the string operator in
region $B$ is related to the correlation matrix $\Gamma_B$ by $\braket{e^{i
    \pi\mathcal N_B}}^2=\det|i\Gamma_B|$. A remaining problem is that
$\lim_{r\to\infty}\det|i\Gamma_B|=0$, which
precludes a numerical evaluation of \eqref{eq:SSB4.5app} on the basis of expressions
\eqref{eq:OK0}. This complication is overcome as follows. We recall
the expression of the determinant of a block matrix 
\be
\det\Bigl|
\begin{pmatrix}
M_{11}&M_{12}\\
M_{21}&M_{22}
\end{pmatrix}
\Bigr|=\det|M_{22}|\det\Bigl|M_{11}-M_{12}M_{22}^{-1}M_{21}\Bigr|\, .
\ee
We then substitute \eqref{eq:Gcal} into \eqref{eq:OK0}, and identify
$2^{\ell+1}\braket{e^{i \pi\mathcal N_B}}^2\{\mathcal G,\mathcal G\}$
and $2^{\ell+1}\braket{e^{i \pi\mathcal N_B}}^2\{\mathcal
G,\overline{\mathcal G}\}$ as the determinants of the matrices
\be\label{eq:auxM}
\begin{pmatrix}
\mathrm I+i\Gamma_{A\cup n}&i\Gamma_{A\cup n,B}\\
i\Gamma_{B,A\cup n}&i\Gamma_B
\end{pmatrix}
\ \text{and}\ \
\begin{pmatrix}
P_n+i\Gamma_{A\cup n}&i\Gamma_{A\cup n,B}\\
i\Gamma_{B,A\cup n}&i\Gamma_B
\end{pmatrix}
\ee
respectively. Rearranging some of the rows and columns we obtain
\be\label{eq:OK1}
\ba
\braket{e^{i \pi\mathcal N_B}}^2\{\mathcal G,\mathcal G\}&=\frac{\det\bigl|\mathrm I_{2\ell}\oplus 0_{2r}\oplus\mathrm I_2+i \Gamma_{A\cup B\cup n}\bigr|}{2^{\ell+1}}\\
\braket{e^{i \pi\mathcal N_B}}^2\{\mathcal G,\overline{\mathcal G}\}&=\frac{\det\bigl|\mathrm I_{2\ell}\oplus 0_{2r}\oplus(-\mathrm I_2)+i \Gamma_{A\cup B\cup n}\bigr|}{2^{\ell+1}}\, .
\ea
\ee
The representations \fr{eq:OK1} are suitable for numerical calculations even in
the limit of large $r$. There is one further simplification: in the
limit $r\to\infty$ we have
\be
\lim_{r\rightarrow\infty}\braket{e^{i \pi\mathcal N_B}}^2\{\mathcal
G,\overline{\mathcal G}\}=-\lim_{r\rightarrow\infty}\braket{e^{i
    \pi\mathcal N_B}}^2\{\mathcal G,\mathcal G\}. 
\label{simplification}
\ee
To see this, we expand the determinants in \eqref{eq:OK1} with respect
to the last $2$-by-$2$ block (from here on we omit the subscript
in $\Gamma_{A\cup B\cup n}$, i.e. $\Gamma\equiv\Gamma_{A\cup B\cup  n}$)  
\begin{multline}\label{eq:sumapp}
\det|\mathrm I_{2\ell}\oplus 0_{2r}\oplus\mathrm I_2+i \Gamma|+\det|\mathrm I_{2\ell}\oplus 0_{2r}\oplus(-\mathrm I_2)+i \Gamma|=\\
2\det\bigl|\Gamma+i\mathrm I_{2\ell}\oplus 0_{2r+2}\bigr|-2\det\bigl|\Gamma_{A\cup B}+i\mathrm I_{2\ell}\oplus 0_{2r}\bigr|\, .
\end{multline}
Using properties of the correlation matrix one could show that the
determinants on the second line approach zero in the limit of large
distance. For the sake of simplicity we propose a different proof,
which is based on the assumption that the limit  
\be\label{eq:existlim}
\lim_{r\rightarrow\infty}\det\bigl|\Gamma_{A\cup B}+i\mathrm I_{2\ell}\oplus 0_{2r}\bigr|
\ee
exists: we demonstrate that the limit cannot be infinite, so the
expression in Eq.~\eqref{eq:sumapp} does tend to zero as
$r\rightarrow\infty$. To this end we consider the $(2\ell+2r)\times
(2\ell+2r)$ correlation matrix $G$ of a generic Gaussian density
matrix, and show that the determinant $\det|G+iI_{2\ell}\otimes
0_{2r}|$ has an upper bound independent of $r$. Hence it cannot
diverge in the limit $r\to\infty$. Our proof is based on the
following facts:
\begin{enumerate}[(a.)]
\item \label{appen:1}$\parallel G\parallel_{op}\leq 1 $, and hence $\parallel G^2\parallel_{op}\leq 1 $ and $\parallel G+i I_{2\ell}\otimes  0_{2r}\parallel_{op}\leq \parallel G\parallel_{op}+1\leq 2$;
\item \label{appen:2}$G+i I_{2\ell}\otimes  0_{2r}$ cannot have more
  than $2\ell$ eigenvalues with absolute values exceeding $1$.
\end{enumerate}
Property~(\ref{appen:1}.) is a consequence of $G$ being the
correlation matrix of a positive semidefinite
Gaussian. Property~(\ref{appen:2}.) can be proved as follows: Let
$\vec w$ a normalized vector with $w_i=0$ for any $i\leq 2\ell$. Then
\be\label{eq:appsm1}
\vec w^\dag (G+i I_{2\ell}\otimes  0_{2r})^\dag (G+i I_{2\ell}\otimes  0_{2r})\vec w=\vec w^\dag G^2\vec w\leq 1\, ,
\ee
where the inequality follows from property~(\ref{appen:1}.). If there
were more than $2\ell$ eigenvalues $\lambda$ of $G+i I_{2\ell}\otimes
0_{2r}$ with modulus larger than $1$, we could find a linear
combination $\vec W=\sum_i c_i\vec v_i$ of the corresponding
normalized eigenvectors $\vec v_i$ with the property $W_i=0$ for any
$i\leq 2\ell$; this leads to a contradiction with \fr{eq:appsm1} since 
\begin{multline}
\sum_{i}c_i^\ast \vec v_i^\dag(G+i I_{2\ell}\otimes  0_{2r})^\dag (G+i I_{2\ell}\otimes  0_{2r})\sum_{j}c_j\vec v_j\\
=\sum_{i}|c_i^2|\lambda_i^2> \sum_{i}|c_i^2|=1\ .
\end{multline}
This completes the proof of property~(\ref{appen:2}.). 

Properties (\ref{appen:1}.) and (\ref{appen:2}.) imply that
\be
|\det|G+i I_{2\ell}\otimes  0_{2r}||\leq 2^{2\ell}\, ,
\ee
which establishes that the determinants in \eqref{eq:sumapp} remain
finite in the limit $r\to\infty$. Concomitantly the expression in
Eq.~\eqref{eq:sumapp} approaches zero as $r\rightarrow\infty$. This
establishes \fr{simplification}. Putting everything together we see that
\eqref{eq:OK1} can be written as
\be\label{eq:SSB5app}
\parallel
\rho_{\ell,o}\parallel_F=\lim_{r\rightarrow\infty}\frac{\sqrt{\det\bigl|\mathrm I_{2\ell}\oplus 0_{2r}\oplus\mathrm I_2+i \Gamma_{A\cup B\cup n}\bigr|}}{2^{\frac{\ell}{2}+1}|m_\perp(t)|}\, ,
\ee
which is Eq.~\eqref{eq:SSB5}. 

We stress that our assumption reagrding the limit~\eqref{eq:existlim}
is equivalent to the existence of the limit in \eqref{eq:SSB5app}. 
From a numerical point of view, this can be inferred from the scaling
analysis of  
\be
\frac{\sqrt{\det\bigl|\mathrm I_{2\ell}\oplus 0_{2r}\oplus\mathrm I_2+i \Gamma_{A\cup B\cup n}\bigr|}}{2^{\frac{\ell}{2}+1}|m_\perp(t)|}\, ,
\ee 
which is still required to check the cluster decomposition hypothesis
(see Fig.~\ref{fig:SSB_norm}).  

The magnetization $|m_\perp(t)|$ can be computed writing a self-consistent equation for Eq.~\eqref{eq:SSB5app} in the case $\ell=1$: From Eq.~\eqref{r1o} we have
\be
\parallel \rho_{1,o}\parallel_F=\sqrt{2}|m_\perp(t)|\, ,
\ee
which together with Eq.~\eqref{eq:SSB5app} gives
\be
4 m_\perp^2(t)=\lim_{r\rightarrow\infty}\sqrt{\det\bigl(\mathrm I_{2}\oplus 0_{2r}\oplus\mathrm I_2+i\Gamma_{\it{1}\cup B\cup n}\bigr)}\, .
\label{final2}
\ee

%%%%%%%%%%%%%%%%%%%%%%%%%%%%%%%%%%%%%%%%%%%%%%%%%%%%%%%%%%%%%%%%%%%%
\section{Conservation laws in spin models with free fermion spectra}
\label{a:cl} %
%%%%%%%%%%%%%%%%%%%%%%%%%%%%%%%%%%%%%%%%%%%%%%%%%%%%%%%%%%%%%%%%%%%%
In this appendix we present a simple construction of the bulk
contribution to local conservation laws of the TFIC on the infinite
line. Our method readily generalizes to other models with free
fermionic spectrum such as the XY chain. Ignoring boundary conditions,
we can use the Jordan-Wigner transformation to express the Hamiltonian
as a quadratic form in Majorana fermions 
\be\label{eq:appH}
H=\frac{1}{2}\sum_{l,n}a_l \mathcal{H}_{l n}a_n\, .
\ee
Here $\mathcal H$ is a skewsymmetric block-circulant matrix
\be\label{eq:H}
\mathcal{H} = \left[
 \begin{array}{ccccc}
\mathcal Y_{0}  & \mathcal Y_{1}   &   \cdots & \mathcal Y_{L-1}  \\
\mathcal Y_{L-1} & \mathcal Y_{0}   & &\vdots\\
\vdots&  & \ddots&\vdots  \\
\mathcal Y_{1}& \cdots  & \cdots  &\mathcal Y_{0}
\end{array}
\right],
\ee
where $\mathcal Y_{n}=-{\mathcal Y}_{L-n}^T$ are 2-by-2 matrices. 
In Fourier space
we have
\be\label{eq:Y}
(\mathcal Y_{n})_{j j^\prime}=\frac{1}{L}\sum_{k=1}^L e^{\frac{2\pi i
    k}{L}n}(Y_k)_{j j^\prime}\, , 
\ee
where $(Y_k)_{jn}=-(Y_{-k})_{nj}$. One can show that a complete set of local
conservation laws is obtained by taking
\be
I_r=\frac{1}{2}\sum_{l,n}a_l \mathcal{I}_{r; l n}a_n\, .
\ee
From Eq.~\eqref{eq:commquad} we see that $[H,I_r]=0$ if and only if
$[\mathcal H,\mathcal I_r]=0$. Similarly one has $[I_r,I_{r^\prime}]=0$ if
and only if $[\mathcal I_r,\mathcal I_{r^\prime}]=0$. Hence the
problem of constructing conservation laws is equivalent
to determining an appropriate set of mutually commuting matrices that
commute with $\mathcal H$. Because the projectors on the
eigenvectors of blocks circulant matrices are block circulant matrices,
we seek $\mathcal I_r$ in block-circulant form 
\be
\mathcal I_r = \left[
 \begin{array}{ccccc}
\bar{\mathcal{Y}}^{(r)}_{0}  & \bar{\mathcal Y}^{(r)}_{1}   &   \cdots & 
\bar{\mathcal Y}^{(r)}_{L-1}  \\
\bar{\mathcal Y}^{(r)}_{L-1} & \bar{\mathcal Y}^{(r)}_{0}   & &\vdots\\
\vdots&  & \ddots&\vdots  \\
\bar{\mathcal Y}^{(r)}_{1}& \cdots  & \cdots  &\bar{\mathcal Y}^{(r)}_{0}
\end{array}
\right].
\ee
Imposing $[\mathcal H,\mathcal I_r]=0$ and $[\mathcal I_r,\mathcal I_{r^\prime}]=0$ we obtain the conditions
\be\label{eq:commY}
[Y_k,\bar{Y}_k^{(r)}]=0\, ,\quad [\bar Y_k^{(r)},\bar{Y}_k^{(r^\prime)}]=0\qquad \forall k\, ,
\ee
where ${\bar Y}_k^{(r)}$ is the Fourier transform~\eqref{eq:Y} of $\bar{\mathcal Y}^{(r)}$.
In the quantum Ising model $Y_k$ are $2$-by-$2$ traceless matrices, so Eq.~\eqref{eq:commY} has the simple solution
\be
\bar{Y}_k^{(r)}=\omega^{(r)}_k \mathrm I+q^{(r)}_k Y_k\, ,
\ee
where $\omega^{(r)}_k=-\omega^{(r)}_{-k}$ and $q^{(r)}_k =q^{(r)}_{-k}
$. Fourier transforming back to position space we have
\be\label{eq:dspace}
\bar{\mathcal Y}^{(r)}_n=\frac{1}{L}\sum_{k=1}^L e^{\frac{2\pi i k}{L}n}\omega_k^{(r)}\mathrm I+\frac{1}{L}\sum_{k=1}^L e^{\frac{2\pi i k}{L}n}q_k^{(r)} Y_k\, .
\ee
We define the `range' of a local conservation as the maximal number of
neighbouring spins involved in its density minus one. 
By construction, the range is equal to the
maximal $|n|$ such that $\bar{\mathcal Y}^{(r)}_n$ is nonzero
(\emph{cf.} Eqs~\eqref{eq:appH}, \eqref{eq:H}).   
For the TFIC one finds that $Y_n=0$ for $|n|>1$, and concomitantly 
the range of the Hamiltonian is $r_{H}=1$. It is straightforward to
identify the conservation laws with ranges $\leq r+1$: from
Eq.~\eqref{eq:dspace} they are such that 
\be
\omega_k=\sum_{n=1}^{r+1}c_n^{-}\sin(n k)\, ,\qquad q_k=\sum_{n=0}^{r+1-r_H}c^+_{n}\cos(n k)\, .
\ee
They can be divided in two classes: one with $q_k=0$, which we denote
by $I^{-}$, and one with $\omega_k=0$, which we denote by
$I^{+}$. Finally, a complete set of conservation laws is given by 
\be\label{eq:conslawF}
\ba
I_r^{+}:&\quad \bar{\mathcal Y}^{+, (r)}_n=\frac{1}{L}\sum_{k=1}^L e^{\frac{2\pi i k}{L}n}\cos(r k) Y_k\\
I_r^{-}:&\quad \bar{\mathcal Y}^{-, (r)}_n=-\frac{2J}{L}\sum_{k=1}^L
e^{\frac{2\pi i k}{L}n}\sin ((r+1) k) \mathrm I\ .
\ea
\ee
These are exactly the conservation laws reported in
Eq.~\eqref{Psit_para}. 

We note that the conservation laws $I_r^{-}$ are independent of the
system details, and can be found in any noninteracting model with a
block circulant structure (see also
Ref.~[\onlinecite{F2:2012}]). Indeed they are originated from the
trivial solution of Eq.~\eqref{eq:commY}, namely the identity. 

%%%%%%%%%%%%%%%%%%%%%%%%%%%%%%%%%
\section{Peculiar aspects of defective GGEs}\label{app:defective}%
%%%%%%%%%%%%%%%%%%%%%%%%%%%%%%%%%

In this appendix we discuss some properties of the defective
generalized Gibbs ensembles defined in Section~\ref{sec:DGGE}. 
We start by recalling the standard variational approach for deriving
statistical ensembles in quantum mechanics. One generally seeks the
density matrix that maximizes the entropy under a given set of
constraints on independent, additive conservation laws $I_j$
\be\label{eq:var}
\delta\btr{}{-\rho\log\rho-\lambda\rho-\sum_j\lambda_j I_j\rho}=0\, .
\ee
The solution of \fr{eq:var} is of the form $\rho\propto
\exp\bigl(\sum_j\lambda_j I_j\bigr)$, which shows that the ensemble
is a function only of the conservation laws appearing in
Eq.~\eqref{eq:var}.  

We now consider the density matrix after a quench. All the ensembles
defined in the main text are compatible with the principle of maximal
entanglement entropy, and the GGE, the truncated GGE, and the
truncated defective GGE can be obtained (\emph{a posteriori}) by means
of the variational approach~\eqref{eq:var}. 

Some complications arise when we consider defective GGEs, in which we
exclude a single integral of motion. From Eq.~\eqref{eq:kappa} we find
that the entanglement entropy density $\sigma_{\rm vN}^{dGGE(+q)}$ of the
defective GGE $\rho_{\rm dGGE}^{(+q)}$\footnote{
This is defined as the limit
$L\rightarrow\infty$ of the finite volume entropy density
$\sigma^{\rm dGGE(+q)}_{\rm
  vN}={\displaystyle\lim_{L\rightarrow\infty}}\frac{1}{L}S_{\rm vN}^{\rm dGGE(+q)}$.
}
is given by 
\be
\sigma_{\rm vN}^{\rm dGGE(+q)}=\int_0^\pi\frac{\mathrm d k}{\pi}H\Bigl(\cos\Delta_k-\kappa_q^+ \frac{\cos(q k)}{\varepsilon(k)}\Bigr)\, ,
\ee
where
$H(x)=-\frac{1+x}{2}\log\frac{1+x}{2}-\frac{1-x}{2}\log\frac{1-x}{2}$. By
writing the defective GGE as in Eq.~\eqref{lambdas}, one can easily
show  that $\frac{\partial \sigma_{\rm vN}^{\rm dGGE(+q)}}{\partial k_q^+}$ is
the Lagrange multiplier associated to the conservation law $I_q^+$
(\emph{cf.} Eq.~\eqref{lagr}): if the maximum of the entanglement
entropy is not at the boundaries of the domain of $k_q^+$, then the
equation $\frac{\partial \sigma_{\rm vN}^{\rm dGGE(+q)}}{\partial k_q^+}=0$
has a solution, and $\rho_{\rm dGGE}^{(+q)}$ can be obtained from
Eq.~\eqref{eq:var}. In the absence of peculiar constraints, one would
expect the maximum to be generally a stationary point of the
entanglement entropy.  
\begin{figure}[ht]
\begin{center}
\includegraphics[width=0.45\textwidth]{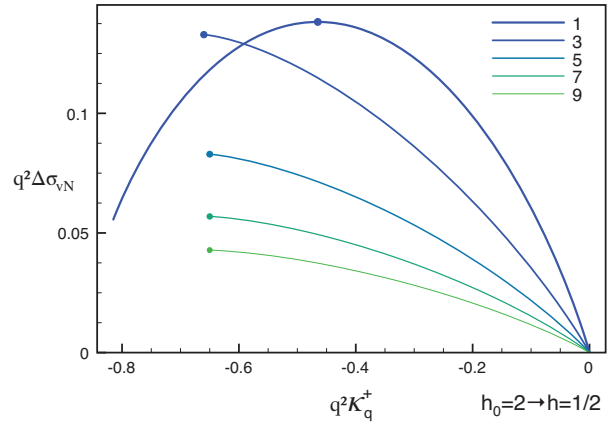}
\caption{The difference of entanglement entropy densities $\Delta\sigma_{vN}=\sigma_{vN}^{dGGE(+q)}-\sigma_{vN}^{GGE}$ as a function of the parameter $\kappa_q^+$ for the same quench shown in Fig.~\ref{fig:local2} (the legend indicates the value of $q$). The points have the maximal entropy and correspond to the lines plotted in Fig.~\ref{fig:local2}. Only for $q=1$ the entanglement entropy is maximal at a stationary point.
}\label{fig:defentropy}
\end{center}
\end{figure}
However, quenches in translationally invariant noninteracting models
are very special since the initial state is a simultaneous  eigenstate
of an infinite number of local conservation laws. This substantially
reduces the degrees of 
freedom, and can result in an exceptionally small domain for $k_q^+$
(which may not include a stationary point).  
In Fig.~\ref{fig:defentropy} we show this paradoxical behaviour for the
same set of parameters used in Fig.~\ref{fig:local2}.  
Besides the pathological cases of even $q$, in which the curves
collapse to the point $\kappa^+_q=0$, the effect of the reduction of
degrees of freedom is reflected in the ``truncated'' shape of
the curves for $q\neq 1$, which turn out to be strictly decreasing
functions of $\kappa_q^+ $.  
The limiting procedure~\eqref{rho_dGGE}  selects the
value of $\kappa_q^+$ corresponding to the maximal entanglement
entropy (the circles in Fig.~\ref{fig:defentropy}).

\end{document}